\documentclass[aps,twocolumn,showpacs,preprintnumbers,amsmath,amssymb]{revtex4}

\usepackage{graphicx}
\usepackage{dcolumn}
\usepackage{bm}
\usepackage{color}
\usepackage{amssymb}   

\begin{document}

\title{Tests of mode-coupling theory in two dimensions}

\author{Fabian Weysser$^1$}
\author{David Hajnal$^2$}
\affiliation{$^1$Fachbereich Physik, Universit\"at Konstanz, D-78457 Konstanz, Germany\\
$^2$Institut f\"ur Physik, Johannes Gutenberg-Universit\"at Mainz, Staudinger Weg 7, D-55099 Mainz, Germany}

\date{\today}

\begin{abstract}
We analyze the glassy dynamics of a binary mixtures of hard disks in two dimensions. Predictions of the Mode-Coupling theory
(MCT) are tested with extensive Brownian dynamics simulations. Measuring the collective particle density correlation functions
in the vicinity of the glass transition we verify four predicted mixing effects. For instance, for large size disparities,
adding a small amount of small particles at fixed packing fraction leads to a speed up in the long time dynamics,
while at small size disparity it leads to a slowing down. Qualitative features of the non-ergodicity parameters and the
$\beta$-relaxation which both depend in a non-trivial way on the mixing ratio are found in the simulated correlators.
Studying one system in detail we are able to determine its ideal MCT glass transition point as $\varphi^c = 0.7948$ and test
MCT predictions quantitatively.
 \end{abstract}


\maketitle

\section{Introduction}

Simulations of simple model systems provide a powerful and important mean to study the interesting phenomenon known as glass transition \cite{KobBinderBuch}.
Considering a binary mixture of hard disks close to vitrification in a computer simulation unveils the opportunity to investigate a system, that is simple
enough so that the particle interactions can be described theoretically, but exhibits the full range of glassy behavior as auto-correlation functions with
two step relaxation processes showing divergent relaxation time scales on approaching the glass transition. 

In 1984, mode couping theory (MCT), which is  a microscopic theory for glassy dynamics, was introduced by Bengtzelius, G\"otze and Sj\"olander and Leutheusser \cite{BengtMCT, leutheusser1984}. It was able to correctly predict many features of the complex dynamics of glass-forming liquids and studied in the subsequent two decades in great detail by G\"otze and coworkers; see Refs.~\cite{goetzeBuch, sjogren1992} for detailed reviews.
The basic version of MCT considers isotropic and homogeneous one-component liquids in
three spatial dimensions (3D). The only model-dependent input is given by the static structure factors of the considered liquid.
The most prominent prediction of MCT is a dynamic transition from a liquid into an ideal nonergodic glassy state at some critical temperature or particle density.

In reality,  one-component (monodisperse) systems do not serve as good glass-formers since they tend to form crystals
rather than amorphous solids. Crystallization can be suppressed by using polydisperse systems. The simplest polydisperse system
is a binary mixture.
Since it is well-known that adding a second component to a one-component liquid may strongly influence
both its static and dynamic properties, G\"otze and Voigtmann \cite{voigtmann2003} have investigated systematically
the glass transition behavior of binary hard spheres in 3D.
They have found four mixing effects:
(i) small size disparities stabilize the glass, (ii) large size disparities stabilize the liquid,
increasing the concentration of the smaller particles leads to both (iii) an increase in the plateau values of the normalized correlation functions for intermediate times for wave numbers that are not too small and (iv) a slowing down of the relaxation of the correlators of the bigger particles towards their plateaus.
These results qualitatively agree with those from dynamic light scattering experiments \cite{Henderson96,Williams01, voigtmann2003-2} and molecular dynamics simulations \cite{foffi2003,foffi2004}.

Several physical phenomena like equilibrium phase transitions strongly depend on the spatial dimensionality $\mathcal{D}$.
Thus, there naturally appears the question about the $\mathcal{D}$-dependence of the glass transition.
From a fundamental point of view, there are interesting studies concerning glass transitions in high dimensions,
see for instance the recent publications of Schmid and Schilling \cite{Schmid10} or Ikeda and Miyazaki \cite{miyazaki2010} and the references therein.
In the present paper, we will restrict ourselves to the case $\mathcal{D}=2$.
An experimental realization of a model glass-former in two dimensions (2D) was presented by
Ebert et al. \cite{ebert2009}. They consider binary mixtures of super-paramagnetic colloidal particles confined at a water-air interface which interact 
via repulsive dipole-potentials. The magnetic moments are induced by an external magnetic 
field perpendicular to the water interface. The results for time-dependent correlation functions measured by video microscopy clearly exhibit slow glassy dynamics as found by K\"onig et al.~\cite{koenig2005} and Mazoyer et al.~\cite{mazoyer2009}.
Computer simulation results of Santen and Krauth \cite{santen2000} for polydisperse hard disks in 2D also give evidence
for the existence of a dynamic glass transition in 2D. Bayer et al.~\cite{bayer2007} have explored the question on the $\mathcal{D}$-dependence
of the glass transition by solving the mode-coupling equations for a one-component system of hard disks in 2D.
They have found an ideal glass transition. On a qualitative level,
the results of Bayer et al.~for the glass transition scenario for monodisperse hard disks in 2D
are very similar to corresponding MCT results of Franosch et al. \cite{franosch1997} for one-component systems  of hard spheres in 3D.

The MCT study of Bayer et al. \cite{bayer2007}
was extended to binary mixtures of hard disks by Hajnal et al. \cite{hajnal2009} and also to binary mixtures of dipolar particles
in 2D \cite{hajnal2010}. For binary hard disks in 2D the same four mixing effects occur as have been reported before by G\"otze and
Voigtmann \cite{voigtmann2003} for binary mixtures of hard spheres in 3D. Furthermore, it was shown that the glass transition diagram
for binary hard disks in 2D
strongly resembles the corresponding random close packing diagram. This fact is a  hint for the applicability of 
the MCT approximations in 2D.

So far, the MCT results of Hajnal et al. \cite{hajnal2009} for the dynamics of binary hard disks in 2D have not yet been tested
systematically within the framework of atomistic computer simulations. To fill in this gap is the main motivation of our contribution.
For this purpose we perform Brownian dynamics (BD) simulations for binary mixtures of hard disks in 2D.
First, we verify the existence of the four mixing effects predicted by MCT which were briefly described above.
Second, we present a quantitative comparison of time-dependent collective density correlators from MCT and our BD simulations.

In order to achieve a self-contained presentation, we have organized the paper as follows: in Sects.~II and III
we introduce the correlation functions of central interest and we define our model system. In Sect.~IV we describe our
BD  simulation techniques. Sect.~V contains a brief review of the equations and central predictions of MCT. We present our results
in Sect.~VI. We summarize and conclude in Sect.~VII.

\section{Preliminaries}

\subsection{Matrix algebra}

\label{matrix_algebra}

In the following, we will make use of the compact mathematical notation introduced in Ref.~\cite{hajnal2009}.
Bold symbols $\boldsymbol{A}$, $\boldsymbol{B}$ etc.
{denote arrays of  $M\times M$ matrices} whose components $\boldsymbol{A}_k$, $\boldsymbol{B}_k$
are labeled by subscript Latin indices.
Their elements $A_k^{\alpha\beta}$, $B_k^{\alpha\beta}$,
also denoted by $(\boldsymbol{A})_k^{\alpha\beta}$, $(\boldsymbol{B})_k^{\alpha\beta}$,
are indicated by superscript Greek indices.
Matrix products are defined component-wise, e.g. $\boldsymbol{C}=\boldsymbol{A}\boldsymbol{B}$ means $\boldsymbol{C}_k=\boldsymbol{A}_k\boldsymbol{B}_k$ for all $k$.
$\boldsymbol{A}$ is called positive-(semi-)definite,
($\boldsymbol{A}\succeq\boldsymbol{0}$) $\boldsymbol{A}\succ\boldsymbol{0}$ if this is true for all $\boldsymbol{A}_k$.
For discretized model systems where $k$ is restricted to a finite number of values, we define the
standard scalar product
$(\boldsymbol{A}|\boldsymbol{B})=\sum_k\sum_{\alpha,\beta}(A_k^{\alpha\beta})^*B_k^{\alpha\beta}$
where the superscript $*$ stands for complex conjugation. The standard norm of $\boldsymbol{A}$ is then given by
$|\boldsymbol{A}|=\sqrt{(\boldsymbol{A}|\boldsymbol{A})}$.

\subsection{Density correlators}

We consider an isotropic and homogeneous classical fluid consisting of $M$ macroscopic components each containing $N_\alpha$ particles 
of a species $\alpha$ in $\mathcal{D}$ spatial dimensions. The total number of particles in the system is then given by $N=\sum_{\alpha=1}^M N_{\alpha}$. 
Let $n^{\alpha}(\vec{r},t)=\sum_{i=1}^{N_{\alpha}}\delta[\vec{r}-\vec{r}_{\alpha,i}(t)]$ denote the
time-dependent microscopic particle density of the component $\alpha$ of the liquid where
$\vec{r}_{\alpha,i}(t)$ is the position of particle $i$ of the component $\alpha$ at time $t$
and $\delta[\cdot]$ is the Dirac delta distribution.
The time-dependent density fluctuation
of the component $\alpha$ of the liquid
at the wave vector $\vec{k}\neq\vec{0}$ is given by the spatial Fourier transform
\begin{equation}
\label{fluctuation}
n_{\vec{k}}^{\alpha}(t)=\sum_{i=1}^{N_{\alpha}}\exp[i\vec{k}\cdot\vec{r}_{\alpha,i}(t)]
\end{equation}
of the particle density $n^{\alpha}(\vec{r},t)$.
We focus our discussion on
the matrix  $\boldsymbol{\Phi}(t)$
of time-dependent partial autocorrelation functions of density fluctuations, which
provide a statistical description of a multicomponent liquid.
For $t\geq0$, its components at wave number $k$ are defined by the expressions
\begin{eqnarray}
\label{Phi}
\Phi_k^{\alpha\beta}(t)&=&\left<N^{-1}[n_{\vec{k}}^{\alpha}(t)]^*n_{\vec{k}}^{\beta}(0)\right>_{TL},\quad k>0,\\
\label{Phi0}
\Phi_0^{\alpha\beta}(t)&=&\lim_{k\rightarrow0^+}\Phi_k^{\alpha\beta}(t),
\end{eqnarray}
where $\left<\dots\right>_{TL}$ stands for canonical averaging followed by carrying out the thermodynamic limit.
The zero-time value of the correlation matrix defines
the normalization $\boldsymbol{\Phi}(0)=\boldsymbol{S} \succ\boldsymbol{0}$, the
positive-definite static structure factor matrix whose elements obey $\lim_{k\rightarrow\infty}S_{k}^{\alpha\beta}=x_{\alpha}\delta_{\alpha\beta}$.
Here $\delta_{\alpha\beta}$ depicts
the Kronecker delta and $x_{\alpha}=N_{\alpha}/N$ the particle number concentration of the component $\alpha$.

\section{Model system}

\label{model_system}

In this work we investigate binary mixtures of hard disks in 2D with diameters $d_{\alpha}$
which are distributed isotropically and homogeneously with total particle number density $n$.
Consisting of ``big'' ($\alpha=b$) and ``small'' ($\alpha=s$) particles with diameters $d_s\leq d_b$, the system
is coupled to a heat bath with thermal energy $k_BT$ and its dynamics is governed by Brownian motion.
The masses $m_{\alpha}$ and the single-particle short-time diffusion coefficients $D_{\alpha}^0$ are set to
$m_{s}=m_b \equiv m_0$ and $D_{s}^0=D_b^0 \equiv D_0$, for simplicity.

It is well-known, that the thermodynamic equilibrium state of the considered model system depends on three
independent control parameters.  Making use of this implicitness we choose 
them to be the total 2D packing fraction $\varphi=n(\pi/4)(x_b d_b^2+x_s d_s^2)$,
the particle number concentration $x_s=N_s/N$ of the smaller disks, and the size ratio $\delta=d_s/d_b$.

\section{Brownian dynamics simulation}

\label{brownian_sim}
The basic concept of the algorithm has been described in detail in
three dimensions in \cite{scala2007}  and can easily be adapted to the 2D model specified above \cite{henrich2009}. 
We consider binary mixtures of hard disks with the size ratios of $\delta = d_s/d_b \in \lbrace 5/7, 1/3 \rbrace$  with particle number concentrations
$x_s \in \lbrace 0.4, 0.5, 0.6, 0.7, 0.8 \rbrace$. 
$N=1000$ hard disks move in a 2D simulation box of volume $V$ 
with periodic boundary conditions at 
packing fraction $\varphi $ as defined in section \ref{model_system}. 
After placing the particles on their initial positions we provide Gaussian distributed velocities
with variance $\langle |{\vec v}_{\alpha,i}|^2\rangle\equiv v^2_0$.
To propagate the system at time $t$ forward in
time, we employ a semi-event-driven algorithm. For every particle, e.g. for particle $i$ of species $\alpha$ at
the time $t$, the algorithm determines the possible collision time
$\Delta t^{\alpha\beta}_{ij}$ with any other particle. This is easily achieved by solving the
equation
\begin{equation}
\frac{d_{\alpha} + d_{\beta}}{2} = |\vec{r}_{ij}^{\, \alpha\beta}+ {\vec v}_{ij}^{\, \alpha\beta} \Delta t^{\alpha\beta}_{ij}|
\end{equation}
where ${\vec r}_{ij}^{\, \alpha\beta}=\vec{r}_{\beta,j}-\vec{r}_{\alpha,i}$ denotes the vector pointing from the center of particle $i$ of species $\alpha$ to
the center of particle $j$ of species $\beta$.
${\vec v}_{ij}^{\, \alpha\beta}=\vec{v}_{\beta,j}-\vec{v}_{\alpha,i}$ denotes the corresponding relative velocity. The smallest solution $\Delta t=\min\{\Delta t^{\alpha\beta}_{ij}\}$
for all particle pairs determines the next event in the algorithm. All
particles can then be propagated with constant velocity according to
${\vec r}_{\alpha,i}(t+ t')= {\vec r}_{\alpha,i}(t)+{\vec v}_{\alpha,i} t'$
for all $t' \in [0,\Delta t]$. At time $t+\Delta t$, for two colliding particles
the elastic binary collision laws impose new velocities 
\begin{equation}
{\vec u}_{\alpha,i}={\vec v}_{\alpha,i}+|\vec{r}_{ij}^{\, \alpha\beta}|^{-2}
(\vec{r}_{ij}^{\, \alpha\beta}\cdot \vec{v}_{ij}^{\, \alpha\beta})\vec{r}_{ij}^{\, \alpha\beta},
\end{equation}
\begin{equation}
{\vec u}_{\beta,j}={\vec v}_{\beta,j}-|\vec{r}_{ij}^{\, \alpha\beta}|^{-2}
(\vec{r}_{ij}^{\, \alpha\beta}\cdot \vec{v}_{ij}^{\, \alpha\beta})\vec{r}_{ij}^{\, \alpha\beta}.
\end{equation}
Due to the boundary conditions any particle in the vicinity of the
box-boundary can collide with an image particle coming from the
other end of the box with the size $L=\sqrt V$.

So far, the algorithm described above yields ballistic motion.
In order to mimic Brownian motion we modify it by introducing
a thermostat which at every integer-multiple of the
time $\tau_B v_0 /d_s=0.01$ triggers a so-called Brownian step.
In the Brownian step, all particle velocities are
freshly drawn from a Gaussian distribution with variance
$m_0v^2_0/(k_BT)=2$ for all particles. 
This assures that the particles move diffusively with a short-time diffusion coefficient $D_0/(v_0d_s)=0.005$
on time scales which are large compared to $\tau_B$.

As the system starts from a cubic lattice it is necessary to wait
for the system to relax before meaningful stationary averages can be
taken. Equilibration was performed with Newtonian dynamics (without imposing the Brownian step)
for $10^5$ time steps in units of $d_s/v_0$. We assume that the system is equilibrated, when the time-dependent correlation functions do not depend on the time origin.
Correlation functions (with imposing the Brownian time step) were measured in a time window of $10^6$ time steps in units of $d_s/v_0$ which is equivalent to
$2551.02$ time steps in units of $d_b^2/D_0$. The collective density correlation functions
given by Eq.~(\ref{Phi}) can directly be calculated from the particle positions.

We selected systems with radius ratios $\delta=5/7$ and $\delta=1/3$. 
Simulations were performed for $x_s \in \lbrace 0.1,\, 0.2, ...,0.9 \rbrace$ and the structure factors and radial distribution functions were compared with the
Percus Yevick results making it possible to exclude the ones with crystallization. For $\delta=5/7$ and $x_s \in \lbrace 0.4, \, 0.5, \, 0.6, \, 0.7 \rbrace$  and for $\delta = 1/3$ and $x_s \in \lbrace 0.5, \, 0.6, \, 0.7, \, 0.8 \rbrace$ we found suitable candidates which are still amorphous at high packing fractions $\varphi$ and thus allow us to investigate the glassy behavior.

\section{Mode-coupling theory}

\subsection{Basic equations}

The mode-coupling theory (MCT) is based on the
exact Zwanzig-Mori equation with a subsequent application of the mode-coupling approximations \cite{goetzeBuch}.
For Brownian dynamics it reads
\begin{equation}
\label{zwanzig_mori}
\boldsymbol{\tau}\boldsymbol{\dot{\Phi}}(t)+\boldsymbol{S}^{-1}\boldsymbol{\Phi}(t)+\int_0^t\mathrm{d} t'
\boldsymbol{m}(t-t')\boldsymbol{\dot{\Phi}}(t')=\boldsymbol{0}.
\end{equation}
The components of the  matrix of microscopic relaxation times $\boldsymbol{\tau}$
shall be approximated by $\tau_k^{\alpha\beta}=\delta_{\alpha\beta}/(k^2D^0_{\alpha}x_{\alpha})$
where $D^0_{\alpha}$ is the single-particle short-time diffusion coefficient of a
tagged particle of species $\alpha$ inside the fluid.
MCT approximates the memory kernel $\boldsymbol{m}(t)$ by a symmetric bilinear functional
\begin{equation}
\boldsymbol{m}(t)=\boldsymbol{\mathcal{F}}[\boldsymbol{\Phi}(t),\boldsymbol{\Phi}(t)].
\end{equation}
For a multicomponent liquid in $\mathcal{D}\geq2$ spatial dimensions it reads \cite{hajnal2009}
\begin{eqnarray}
\label{memory}
\mathcal{F}_k^{\alpha\beta}[\boldsymbol{X},\boldsymbol{Y}]&=&\frac{\Omega_{\mathcal{D}-1}}{(4\pi)^\mathcal{D}}
\sum_{\alpha',\beta',\alpha'',\beta''}\int_0^\infty \mathrm{d} p\int_{|k-p|}^{k+p} \mathrm{d} q\nonumber\\
&&\times V^{\alpha\beta;\alpha'\beta',\alpha''\beta''}_{k;p,q}X_p^{\alpha'\beta'}Y_q^{\alpha''\beta''}
\label{bilinear}
\end{eqnarray}
where the so-called vertices are given by
\begin{equation}
\label{vertex_1}
V^{\alpha\beta;\alpha'\beta',\alpha''\beta''}_{k;p,q}=
\frac{n}{x_{\alpha}x_{\beta}}\frac{pq}{k^{\mathcal{D}+2}}v_{kpq}^{\alpha\alpha'\alpha''}v_{kpq}^{\beta\beta'\beta''},
\end{equation}
\begin{equation}
\label{vertex}
v^{\alpha\beta\gamma}_{kpq}=\frac{(k^2+p^2-q^2)c_p^{\alpha\beta}\delta_{\alpha\gamma}
+(k^2-p^2+q^2)c_q^{\alpha\gamma}\delta_{\alpha\beta}}{[
4k^2p^2-(k^2+p^2-q^2)^2]^{(3-\mathcal{D})/4}}.
\end{equation}
$c_k^{\alpha\beta}$ denote the direct correlation functions
and $\Omega_{\mathcal{D}}=2{\pi}^{\mathcal{D}/2}/\Gamma(\mathcal{D}/2)$ is the surface of the $\mathcal{D}$-dimensional unit sphere. $\Gamma(x)$ is
the gamma function.
$\boldsymbol{c}$ is related to $\boldsymbol{S}$ via the Ornstein-Zernike equation
$(\boldsymbol{S}^{-1})_k^{\alpha\beta}=\delta_{\alpha\beta}/x_{\alpha}-nc_k^{\alpha\beta}$.

\subsection{Discretized model}
For practical purposes, we follow Ref.~\cite{hajnal2009} and discretize the wave number  $k$ to a
finite, equally spaced grid of $K$ points $k=(\hat{o}_d+\hat{k})\Delta k$ with $\hat{k}=0,1,\dots,K-1$ and $0<\hat{o}_d<1$.
The integrals in Eq.~(\ref{memory}) are then replaced by Riemann sums
\begin{equation}
\label{riemann}
\int_0^\infty \mathrm{d} p\int_{|k-p|}^{k+p} \mathrm{d} q \dots \mapsto (\Delta k)^2 \sum_{\hat{p}=0}^{K-1}  \sum_{\hat{q}=|\hat{k}-\hat{p}|}^{\min\{K-1,\hat{k}+\hat{p}\}}\dots
\end{equation}
and Eq.~(\ref{zwanzig_mori}) represents a finite number of coupled nonlinear integro-differential equations.

The discretized MCT equations shall be applied to the binary hard disk mixture in 2D introduced in Sect.~\ref{model_system}.
For this, we choose  $K=250$, $\Delta k \cdot d_b=0.3$ and $\hat{o}_2=0.303$.
Static input for the MCT calculations presented in this paper is taken from our BD simulations.
For the numerical solution of Eq.~(\ref{zwanzig_mori}) we use the algorithm described in \cite{fuchs1991}.
Our time grids consist of $256$ points and as initial time step size we choose $2.5\cdot10^{-11}$ units of $d_b^2/D_0$.

\subsection{Glass transition singularities}

The non-ergodicity parameters (NEPs) are the elements of the matrix
$\boldsymbol{F}=\lim_{t\rightarrow\infty}\boldsymbol{\Phi}(t)$.
For the discretized model above, it can
be proved \cite{franosch2002} that $\boldsymbol{F}$ is 
(with respect to $\succeq$) the maximum real, symmetric fixed point of
the nonlinear map
\begin{equation}
\label{fixed_point}
\boldsymbol{\mathcal{I}}[\boldsymbol{X}]=\boldsymbol{S}-(\boldsymbol{S}^{-1}+\boldsymbol{\mathcal{F}}[\boldsymbol{X},\boldsymbol{X}])^{-1}.
\end{equation}
For the binary hard disk model, MCT predicts that the liquid-to-glass transition takes place at a locally smooth critical surface $\mathcal{H}$
which we can represent as the critical packing fraction $\varphi^c(x_s,\delta)$ as a function of the concentration of
the smaller disks and the size ratio. At this surface $\boldsymbol{F}$ jumps from $\boldsymbol{0}$ (liquid) to some
$\boldsymbol{F}^c\succ\boldsymbol{0}$ (glass).
Quantities corresponding to critical points shall be indicated by a superscript $c$ in the following.
General properties of $\mathcal{H}$ have been discussed in detail in Ref.~\cite{hajnal2009}.

Besides $\boldsymbol{F}^c$, there are further important quantities characterizing a generic liquid-to-glass transition point
which is an $A_2$ singularity according to the classification of Arnol'd \cite{Arnold}.
Linearization of $\boldsymbol{\mathcal{I}}^c$ around $\boldsymbol{F}^c$ yields
a so-called positive linear map \cite{franosch2002}
\begin{equation}
\label{stability}
\boldsymbol{\mathcal{C}}^c[\boldsymbol{Y}]=2(\boldsymbol{S}^c-\boldsymbol{F}^c)\boldsymbol{\mathcal{F}}^c[\boldsymbol{F}^c,\boldsymbol{Y}](\boldsymbol{S}^c
-\boldsymbol{F}^c)
\end{equation}
with $\boldsymbol{\mathcal{C}}^c[\boldsymbol{Y}]\succeq\boldsymbol{0}$ for all $\boldsymbol{Y}\succeq\boldsymbol{0}$.
This map has a non-degenerated maximum eigenvalue $r=1$ with a corresponding (right) eigenvector $\boldsymbol{H}^c$
and a corresponding left eigenvector $\hat{\boldsymbol{H}}^c$ which is an eigenvector to eigenvalue $r^*=1$ of the
adjoint map of $\boldsymbol{\mathcal{C}}^c$ with respect to the scalar product defined in Sect.~\ref{matrix_algebra}.
These two eigenvectors
are determined uniquely if we require the normalization
\begin{equation}
\label{norm}
(\hat{\boldsymbol{H}}^c|\boldsymbol{H}^c)=(\hat{\boldsymbol{H}}^c|\boldsymbol{H}^c\{\boldsymbol{S}^c-\boldsymbol{F}^c\}^{-1}\boldsymbol{H}^c)=1.
\end{equation}

A further important quantity is the so-called exponent parameter
\begin{equation}
\label{lambda}
\lambda^c=(\hat{\boldsymbol{H}}^c|\{\boldsymbol{S}^c-\boldsymbol{F}^c\} \boldsymbol{\mathcal{F}}^c
[\boldsymbol{H}^c,\boldsymbol{H}^c]\{\boldsymbol{S}^c-\boldsymbol{F}^c\})
\end{equation}
whose value determines the exponents in the asymptotic scaling-laws (see the next section).
These positive exponents are the critical exponent $a$ obeying the relation
\begin{equation}
\label{exp_a}
{\Gamma^2(1-a)}/{\Gamma(1-2a)}=\lambda^c,
\end{equation}
the von Schweidler exponent $b$ satisfying
\begin{equation}
\label{exp_b}
{\Gamma^2(1+b)}/{\Gamma(1+2b)}=\lambda^c,
\end{equation}
and the exponent
\begin{equation}
\label{exp_gamma}
\gamma={(a+b)}/{(2ab)}
\end{equation}
describing the divergence of the time scale for the final relaxation of $\boldsymbol{\Phi}(t)$
to $\boldsymbol{0}$ upon increasing $\varphi$ towards its critical value $\varphi^c$.

\subsection{Asymptotic scaling-laws}

\label{sect_scaling}

Close to the liquid-glass transition, MCT makes universal predictions for the relaxation behavior
of $\boldsymbol{\Phi}(t)$ which can be studied in the framework of asymptotic expansions.
For the following, let us fix $x_s$ and $\delta$ to some specific value and define the
distance parameter
\begin{equation}
\label{epsilon}
\varepsilon=(\varphi-\varphi^c)/\varphi^c.
\end{equation}
Finally, we introduce the separation parameter $\sigma(\varepsilon)$ which is a linear function of $\varepsilon$.
It follows from
\begin{eqnarray}
\label{sigma}
\tilde{\sigma}(\varphi)&=&(\hat{\boldsymbol{H}}^c|\{\boldsymbol{S}^c-\boldsymbol{F}^c\}{\boldsymbol{S}^c}^{-1}
\{\boldsymbol{S}\boldsymbol{\mathcal{F}}[\boldsymbol{F}^c,\boldsymbol{F}^c](\boldsymbol{S}-\boldsymbol{F}^c)\nonumber\\
&&-\boldsymbol{S}^c\boldsymbol{\mathcal{F}}^c[\boldsymbol{F}^c,\boldsymbol{F}^c](\boldsymbol{S}^c-\boldsymbol{F}^c)\})
\end{eqnarray}
by expanding around $\varphi^c$ up to linear order in $\varepsilon$ \cite{goetzeBuch,Voigtmann_PhD}.

\subsubsection{The first scaling-law regime}

For small separation parameters, $\boldsymbol{\Phi}(t)$ develops power-law dynamics located around $\boldsymbol{F}^c$
which is also called the $\beta$-relaxation process.
For times within the so-called first scaling-law regime
defined by $|\boldsymbol{\Phi}(t)-\boldsymbol{F}^c|\ll 1$, also called the $\beta$-scaling regime,
there holds the factorization
theorem \cite{goetzeBuch,franosch1997}
\begin{equation}
\label{factorization}
\boldsymbol{\Phi}(t)-\boldsymbol{F}^c=\boldsymbol{H}^c\mathcal{G}(t)+\mathcal{O}(|\sigma|)
\end{equation}
with the $\beta$-correlator $\mathcal{G}(t)=\mathcal{O}(|\sigma|^{1/2})$ obeying the equation of motion
\begin{equation}
\label{beta_scaling_eq}
\sigma+\lambda^c\mathcal{G}^2(t)=\frac{\mathrm{d}}{\mathrm{d}t}\int_0^t\mathrm{d}t'\mathcal{G}(t-t')\mathcal{G}(t')
\end{equation}
with the divergent initial condition
\begin{equation}
\label{inital_beta}
\mathcal{G}(t\rightarrow 0)=(t/t_0)^{-a}.
\end{equation}
The $\varepsilon$-independent time scale $t_0$ has to be matched to the full solution of Eq.~(\ref{zwanzig_mori})
at $\varepsilon=0$ since in this case the power law occurring in Eq.~(\ref{inital_beta}) is a special solution of
Eq.~(\ref{beta_scaling_eq}) which describes the relaxation of $\boldsymbol{\Phi}(t)$ towards $\boldsymbol{F}^c$.
It is easy to verify that $\mathcal{G}(t)$ obeys the scaling-law
\begin{equation}
\label{scaling_g}
\mathcal{G}(t,\sigma\gtrless 0)=|\sigma|^{1/2}\tilde{\mathcal{G}}(\tilde{t}=t/t_{\sigma},\tilde{\sigma}=\pm 1),
\end{equation}
\begin{equation}
\label{tau_sigma}
t_{\sigma}=t_0|\sigma|^{-\frac{1}{2a}}.
\end{equation}
The master function $\tilde{\mathcal{G}}(\tilde{t})$ obeys Eq.~(\ref{beta_scaling_eq}) with the replacement $\sigma\mapsto \sigma/|\sigma|=\pm 1$
and the initial condition $\tilde{\mathcal{G}}(\tilde{t}\rightarrow 0)=(\tilde{t})^{-a}.$
Eq.~(\ref{beta_scaling_eq}) can be solved by asymptotic series expansions \cite{hajnal2009-2}.
One finds the leading long-time asymptotes
\begin{equation}
\label{beta_long_time_1}
\mathcal{G}(t\rightarrow\infty,\sigma\geq0)=\sqrt{\sigma/(1-\lambda^c)},
\end{equation}
\begin{equation}
\label{beta_long_time_2}
\mathcal{G}(t\rightarrow\infty,\sigma<0)=-(t/\tau)^{b}.
\end{equation}
Eq.~(\ref{beta_long_time_1}) describes the asymptotic behavior of $\boldsymbol{F}$ in the glassy regime
close to $\varphi^c$.
Eq.~(\ref{beta_long_time_2}) is referred to as the von Schweidler law and
describes the initial part of the relaxation of $\boldsymbol{\Phi}(t)$ from
$\boldsymbol{F}^c$ to $\boldsymbol{0}$. Eqs.~(\ref{scaling_g}) and (\ref{tau_sigma})
imply
\begin{equation}
\label{tau_alpha}
\tau=\tilde{\tau}t_0|\sigma|^{-\gamma}
\end{equation}
with the $\varepsilon$-independent constant $\tilde{\tau}$ which demonstrates that
the arrest of $\boldsymbol{\Phi}(t)$ to $\boldsymbol{F}^c$ at $\varphi=\varphi^c$ is
caused by a power-law divergence of the times scale 
for the inset of the relaxation of $\boldsymbol{\Phi}(t)$ from $\boldsymbol{F}^c$ to $\boldsymbol{0}$
upon increasing $\varphi$ towards its critical value $\varphi^c$.

\subsubsection{The second scaling-law regime}

Now we turn to the relaxation process of $\boldsymbol{\Phi}(t)$ from $\boldsymbol{F}^c$ to $\boldsymbol{0}$
within the liquid regime which is also called the $\alpha$-relaxation process.
Considering the limits $\varepsilon\rightarrow 0^-$ and $t\rightarrow \infty$ with fixed $\tilde{t}=t/\tau$, one arrives at the so-called
$\alpha$-scaling law \cite{goetzeBuch,franosch1997}
\begin{equation}
\label{alpha_scaling_law}
\boldsymbol{\Phi}(t)=\tilde{\boldsymbol{\Phi}}^c(\tilde{t})+\mathcal{O}(|\sigma|)
\end{equation}
where the $\varepsilon$-independent master function $\tilde{\boldsymbol{\Phi}}^c(\tilde{t})$ obeys the
equation of motion
\begin{equation}
\label{alpha_scaling_eq}
(\boldsymbol{S}^c)^{-1}\tilde{\boldsymbol{\Phi}}^c(\tilde{t})=\tilde{\boldsymbol{m}}^c(\tilde{t})\boldsymbol{S}^c
-\frac{\mathrm{d}}{\mathrm{d}\tilde{t}}\int_0^{\tilde{t}}
\mathrm{d} \tilde{t}'
\tilde{\boldsymbol{m}}^c(\tilde{t}-\tilde{t}')\tilde{\boldsymbol{\Phi}}^c(\tilde{t}')
\end{equation}
with the memory kernel
\begin{equation}
\tilde{\boldsymbol{m}}^c(\tilde{t})=\boldsymbol{\mathcal{F}}^c[\tilde{\boldsymbol{\Phi}}^c(\tilde{t}),\tilde{\boldsymbol{\Phi}}^c(\tilde{t})]
\end{equation}
and the von Schweidler law
\begin{equation} \label{vsmix}
\tilde{\boldsymbol{\Phi}}^c(\tilde{t}\rightarrow 0)=\boldsymbol{F}^c-(\tilde{t})^b\boldsymbol{H}^c
\end{equation}
as initial condition.
Eq.~(\ref{alpha_scaling_law}) is also called superposition principle due to the following implication:
for times within the so-called second scaling-law regime given by $t\gg\tau$, density correlators
$\Phi_k^{\alpha\beta}(t)$ corresponding to different values of $\varepsilon$ collapse onto master curves
$(\tilde{\boldsymbol{\Phi}}^c)_k^{\alpha\beta}(\tilde{t})$
when they are plotted as functions of $\tilde{t}=t/\tau$.

\subsection{Mixing effects}

As already briefly mentioned in the introduction, MCT predicts for binary hard disks in 2D the existence of
four mixing effects \cite{hajnal2009}. Let us recapitulate these effects in more detail.
\begin{enumerate}
\item[(i)]
For small size disparities the glassy regime is enhanced. For $0.65\lesssim\delta<1$ it is
$\varphi^c(x_s,\delta)<\varphi^c_0$ if $0<x_s<1$, where $\varphi^c_0$ denotes the critical packing
fraction for monodisperse hard disks in 2D. For $\delta=5/7$, for instance, $\varphi^c(x_s)$ develops a single minimum located
at $x_s\approx0.5$.
\item[(ii)]
For larger size disparities the liquid state is stabilized, i.e. for $\delta\lesssim0.37$ it is 
$\varphi^c(x_s,\delta)>\varphi^c_0$ if $0<x_s<1$.
This effect is also called plasticization.
For $\delta=1/3$, for instance, $\varphi^c(x_s)$ develops a single maximum located at $x_s\approx0.85$.
\item[(iii)]
Upon increasing the concentration $x_s$ of the smaller particles the NEPs, and thus also
the plateau values of the normalized correlation functions $\Phi_k^{\alpha\alpha}(t)/S_k^{\alpha\alpha}$ in the liquid regime
for intermediate times, increase for not too small $k$ and all $x_s$.
\item[(iv)]Starting with $x_s=0$ and increasing
the concentration $x_s$ of the smaller particles leads for not too large $k$ to
a slowing down of the
relaxation of the normalized correlators $\Phi_k^{bb}(t)/S_k^{bb}$ of the big particles
towards their plateaus 
in the sense that the $\Phi_k^{bb}(t)/S_k^{bb}$ versus $\log_{10}(t)$ curve becomes
flatter upon increasing $x_s$.
\end{enumerate}
The amplitude of the predicted effects (i) and (ii) is quite small, the total variation of $\varphi^c(x_s,\delta)$ is of the order of $1\%$.
Nevertheless, these small changes in $\varphi^c$ may have a strong influence on time-dependent quantities which are
accessible to our BD simulations. Let us fix some $\delta$ and $\varphi$ such that for all $0<x_s<1$ the condition
$0<[\varphi^c(x_s,\delta)-\varphi]\ll1$ is satisfied.
Eqs.~(\ref{tau_alpha}) and (\ref{alpha_scaling_law}) predict then a strong variation in the time scales $\tau$
for the $\alpha$-relaxation of the correlators $\Phi_k^{\alpha\beta}(t)$. It has been demonstrated \cite{hajnal2009} that an occurring minimum (maximum)
in $\varphi^c(x_s)$ is then directly reflected by an occurring maximum (minimum) in $\tau$.
Using this information, we demonstrate in the following that the mixing effects (i)-(iv) predicted by MCT are indeed 
observable in our BD simulation data.

\section{Results and discussion}

We choose to use a Brownian dynamics simulation for our comparison with MCT results for the following reason:
the behavior of the relaxation times of the system on approaching the glass transition point, the details of the $\alpha$-relaxation,
as well as the NEPs are independent of the microscopic dynamics of supercooled liquids, as Gleim et. al. showed in \cite{gleim1998}.
Nevertheless, the relaxation onto the plateaus is expected to be described better within MCT for Brownian dynamics \cite{gleim1998}.

\subsection{Statics}

\begin{figure}
\includegraphics[width=1\columnwidth]{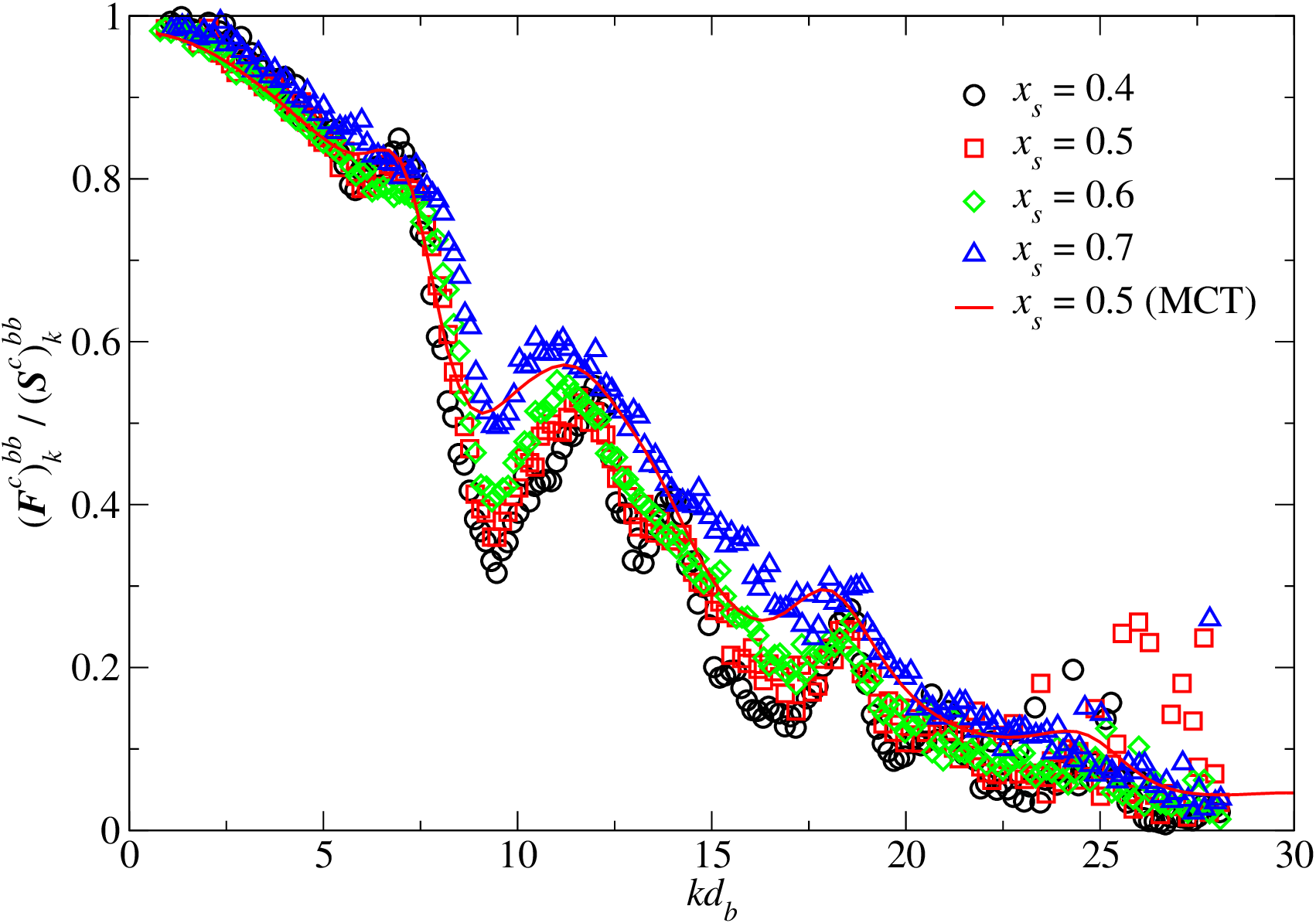}
\caption{(Color online) Normalized critical non-ergodicity parameters of the simulated collective density correlators $\Phi_k^{bb}(t)$ of the big particles,
extracted from Kohlrausch fits for the size ratio $\delta = 5/7$ and the packing fraction $\varphi=0.79$. The number concentrations of small particles $x_s$ vary as labeled in the legend. The solid red line gives the MCT results calculated with a simulated structure factor input at $\varphi^c_{MCT}=0.6920$ and $x_s=0.5$.}
\label{delta1.4-fq-big}
\end{figure}
\begin{figure}
\includegraphics[width=1\columnwidth]{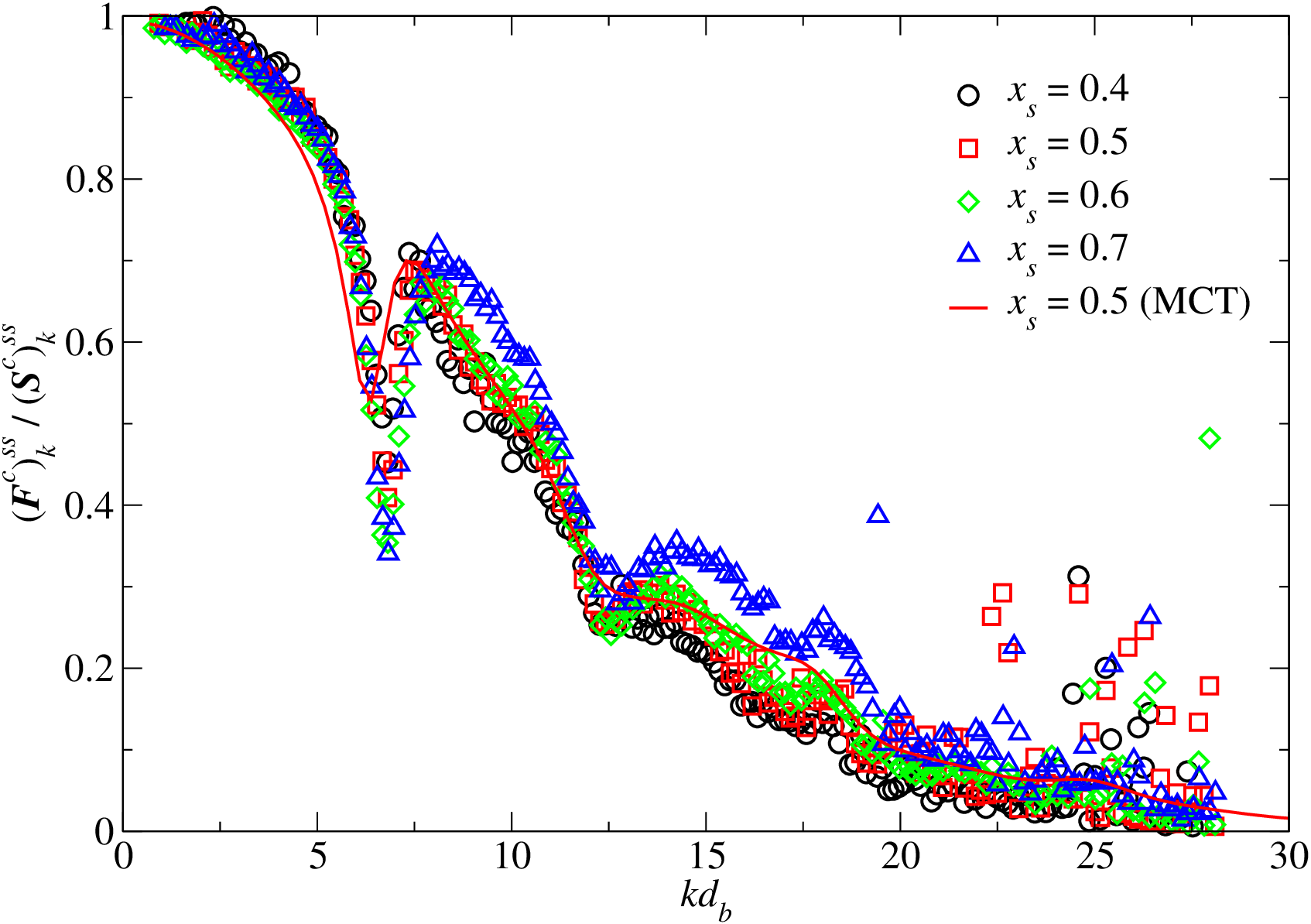}
\caption{(Color online) Normalized critical non-ergodicity parameters of the simulated collective density correlators $\Phi_k^{ss}(t)$ of the small particles,
extracted from Kohlrausch fits for the size ratio $\delta = 5/7$ and the packing fraction $\varphi=0.79$. The number concentrations of small particles $x_s$ are the same as in Fig.~\ref{delta1.4-fq-big}. The solid red line shows the MCT results calculated with simulated structure factors as in Fig.~\ref{delta1.4-fq-big}.}
\label{delta1.4-fq_small.eps}
\end{figure}
The first point we want to address are the non-ergodicity parameters. A common description of the $\alpha$-relaxation is in terms of stretched exponential Kohlrausch laws,
\begin{equation}
{\Phi}_k^{\alpha \alpha}(t) = {A}_k^{\alpha \alpha} \exp \left[-(t/\tilde{\tau}_k^{\alpha \alpha})^{\beta_k^{\alpha \alpha}} \right],
\end{equation}
with the stretching exponent $\beta_k^{\alpha \alpha}$, a relaxation time scale ${\tilde{\tau}}_k^{\alpha \alpha}$ and the amplitude $A_k^{\alpha \alpha}$. 
For structural relaxation in equilibrium system $\beta_k^{\alpha \alpha} <1$ is required. The $\alpha$-master function from MCT equation~\eqref{alpha_scaling_eq} is different from the Kohlrausch form, however the theory predicts that for large wave numbers the two functional forms become identical and $\beta_k^{\alpha \alpha} \to b$ \cite{fuchs1994}.
The Kohlrausch amplitude $A_k^{\alpha \alpha}$ provides an estimate for the MCT NEPs $({\bm F }^c)_k^{\alpha \alpha}$. Since the $\alpha$-process starts below this plateau value, $ A_k^{\alpha \alpha} \le ({\bm F }^c)_k^{\alpha \alpha}$ should hold.  However, in practice the separation of the $\alpha$-process from the $\beta$-relaxation is not clear enough to fulfill this prediction.

Kohlrausch fits are hindered by some subtle problems: lacking a clear separation of the $\alpha$-process, the fit parameters inclose a dependence on the fit range. A priori it is unclear how to choose the optimal fit range,
as for very long times one expects the relaxation to become (non stretched) exponential again, and for short times, deviations stemming from the $\beta$-relaxation hamper the choice. The fit range was fixed, so that the parameters only exhibit the weakest (the region were they are almost constant) dependence on the boundaries.
This procedure leads to $t \in  [7.65,2551.02]$ for $\delta=5/7$ and $\varphi = 0.79$ with the various $x_s$. For $\delta=1/3$ and $\varphi = 0.81$ we used $t \in  [1.39,555.56]$ for the various $x_s$.

Fig.~\ref{delta1.4-fq-big} shows approximate values for the normalized critical NEPs $(\boldsymbol{F}^c)_k^{bb}/(\boldsymbol{S}^c)_k^{bb}$ for the big disks at $\delta=5/7$ and different values for $x_s$, extracted from our BD simulation data via Kohlrausch-fits.
Corresponding results for the smaller disks are shown in Fig.~\ref{delta1.4-fq_small.eps}.
In both Fig.~\ref{delta1.4-fq-big} and Fig.~\ref{delta1.4-fq_small.eps}, we have also included critical NEPs for $x_s=0.5$, calculated via MCT with BD-simulated structure factors as input, with MCT yielding a critical packing fraction of $\varphi_{MCT}^c=0.6920$ for the simulated input.
On a qualitative level, our MCT results are in good agreement with our BD simulation results.  For the big particles the relation ${A}_k^{\alpha \alpha}  \le  ({\bm F }^c)_k^{\alpha \alpha}$ is well fulfilled for all $kd_b$ except for some outliers. The same holds for the small particles but for $kd_b \lesssim 5$ the Kohlrausch fit yields smaller estimations for the NEPs.  In both Fig.~\ref{delta1.4-fq-big} and Fig.~\ref{delta1.4-fq_small.eps},
we observe a slight increase in the BD simulation results for the NEPs upon increasing $x_s$ which is on a qualitative level in agreement with previous MCT results \cite{voigtmann2003,hajnal2009}.

\begin{figure}
\includegraphics[width=1\columnwidth]{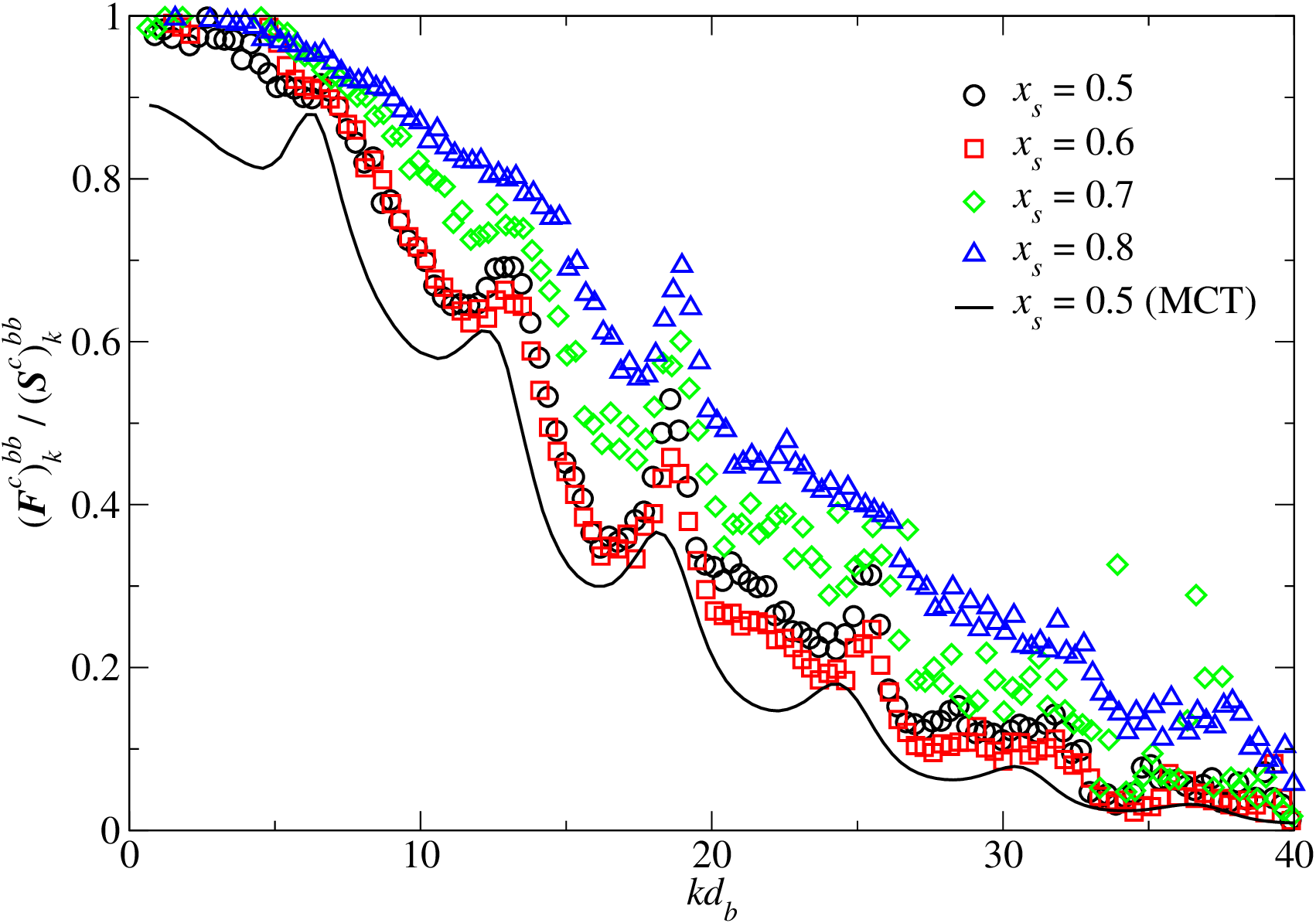}
\caption{(Color online) Normalized critical non-ergodicity parameters of the simulated collective density correlators $\Phi_k^{bb}(t)$ of the big particles,
extracted from Kohlrausch fits for the size ratio $\delta = 1/3$ and the packing fraction $\varphi=0.81$. The various number concentrations $x_s$ are as depicted in the legend. MCT results using simulated structure factors as input with $\varphi_{MCT}^c=0.6991$ are shown as black solid line for $x_s=0.5$.}
\label{delta3-fq-big}
\end{figure}
\begin{figure}
\includegraphics[width=1\columnwidth]{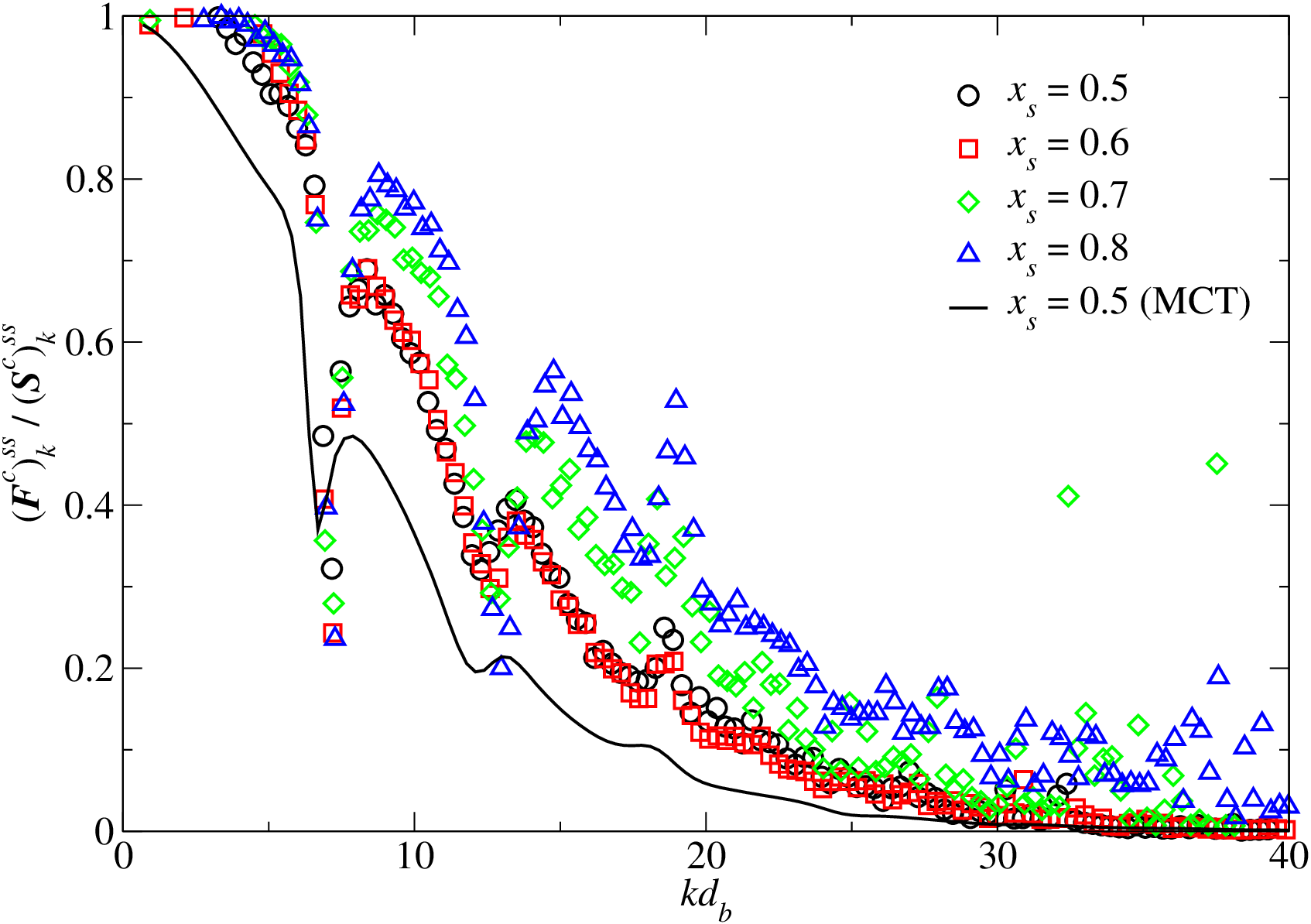}
\caption{(Color online) Normalized critical non-ergodicity parameters of the simulated collective density correlators $\Phi_k^{ss}(t)$ of the small particles,
extracted from Kohlrausch fits for the size ratio $\delta = 1/3$ and the packing fraction $\varphi=0.81$. The color and symbol coding for the different $x_s$ is the same as in Fig.~\ref{delta3-fq-big}. The solid black line shows the MCT results calculated with simulated structure factors as used in Fig.~\ref{delta3-fq-big}.}
\label{delta3-fq_small.eps}
\end{figure}

Fig.~\ref{delta3-fq-big} shows approximate values for the normalized critical NEPs $(\boldsymbol{F}^c)_k^{bb}/(\boldsymbol{S}^c)_k^{bb}$
for the big disks at $\delta=1/3$ and different values for $x_s$, extracted from our BD simulation data via Kohlrausch-fits.
Corresponding results for the NEPs of the smaller disks are shown in Fig.~\ref{delta3-fq_small.eps}.
In both Fig.~\ref{delta3-fq-big} and Fig.~\ref{delta3-fq_small.eps}, we have also included critical NEPs for $x_s=0.5$,
calculated via MCT with BD-simulated structure factors as input, giving a critical packing fraction of $\varphi_{MCT}^c=0.6991$ for the simulated input. For the present value of $\delta$, we observe that our MCT
calculations yield systematically smaller values for the NEPs, compared to our BD simulations results.
The underestimation of the NEPs may be attributed to the underestimation of $\varphi^c$:
MCT predicts arrest at lower densities, but the NEPs may increase with density as the denser glass is stiffer with respect
to density fluctuations.
The BD simulation  results in both Fig.~\ref{delta3-fq-big} and Fig.~\ref{delta3-fq_small.eps} indicate a systematic
increase in the NEPs upon increasing $x_s$ which is, as expected, more strongly pronounced than for the case $\delta=5/7$
which is on a qualitative level in agreement with previous MCT results \cite{voigtmann2003,hajnal2009}.

We can conclude here, that at least on a qualitative level, our BD simulation results confirm
the $x_s$ and $\delta$ dependences of the NEPs predicted by MCT \cite{voigtmann2003,hajnal2009}.
In particular, we have clearly verified the existence of mixing effect (iii).

\begin{figure}
\includegraphics[width=1\columnwidth]{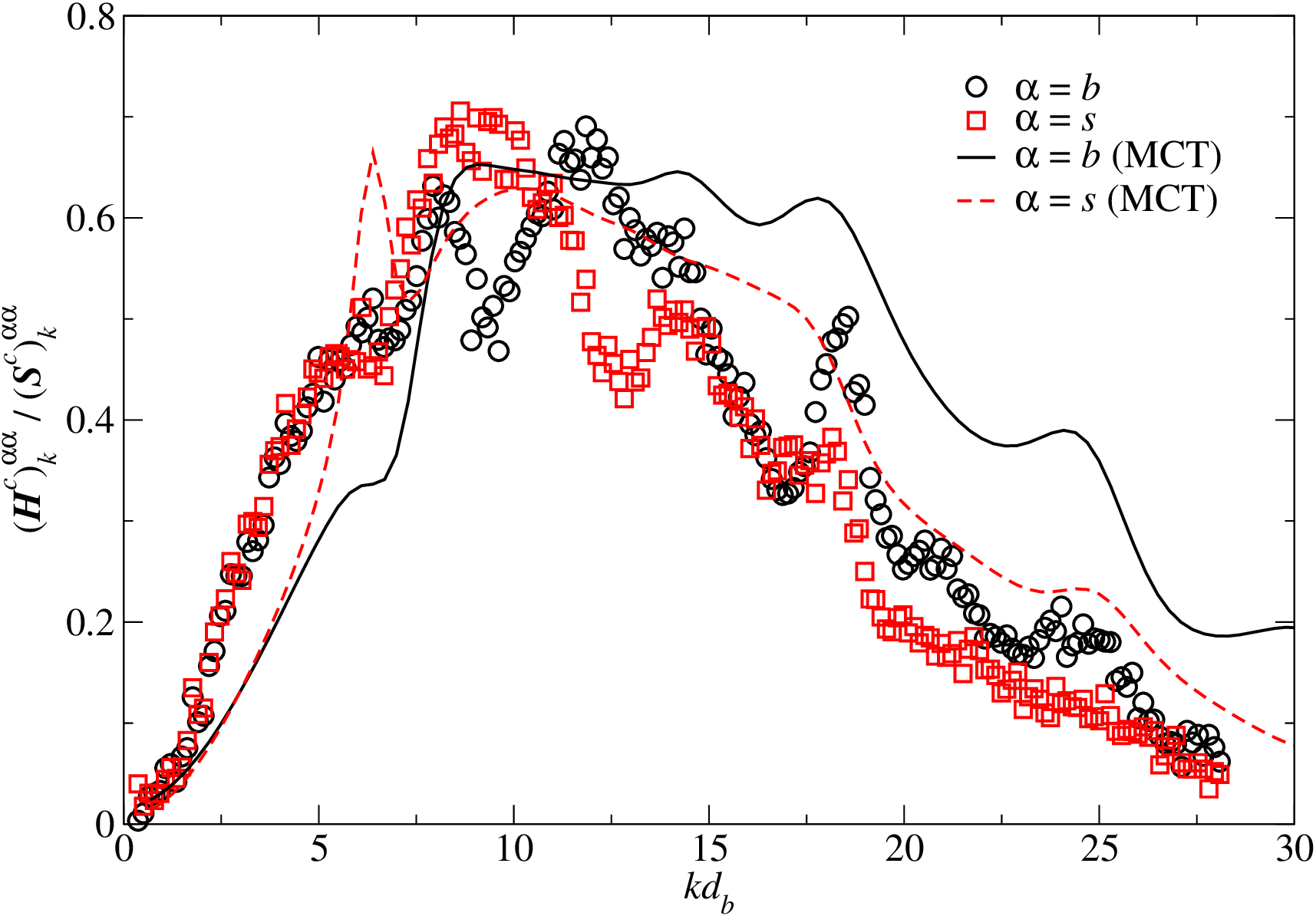}
\caption{(Color online) Critical amplitudes obtained by Eq.~(\ref{Yfunc}) for the big and small particles with $k_0d_b = 2.5$.
The data were extracted from the collective correlators at $\varphi=0.79$, $\delta = 5/7$ and $x_s=0.5$. Solid black and dashed red lines depict the MCT results obtained with the same simulated structure factors as in   Fig.~\ref{delta1.4-fq-big} and Fig.~\ref{delta1.4-fq_small.eps}}
\label{hq-fig_1.4}
\end{figure}
\begin{figure}
\includegraphics[width=1\columnwidth]{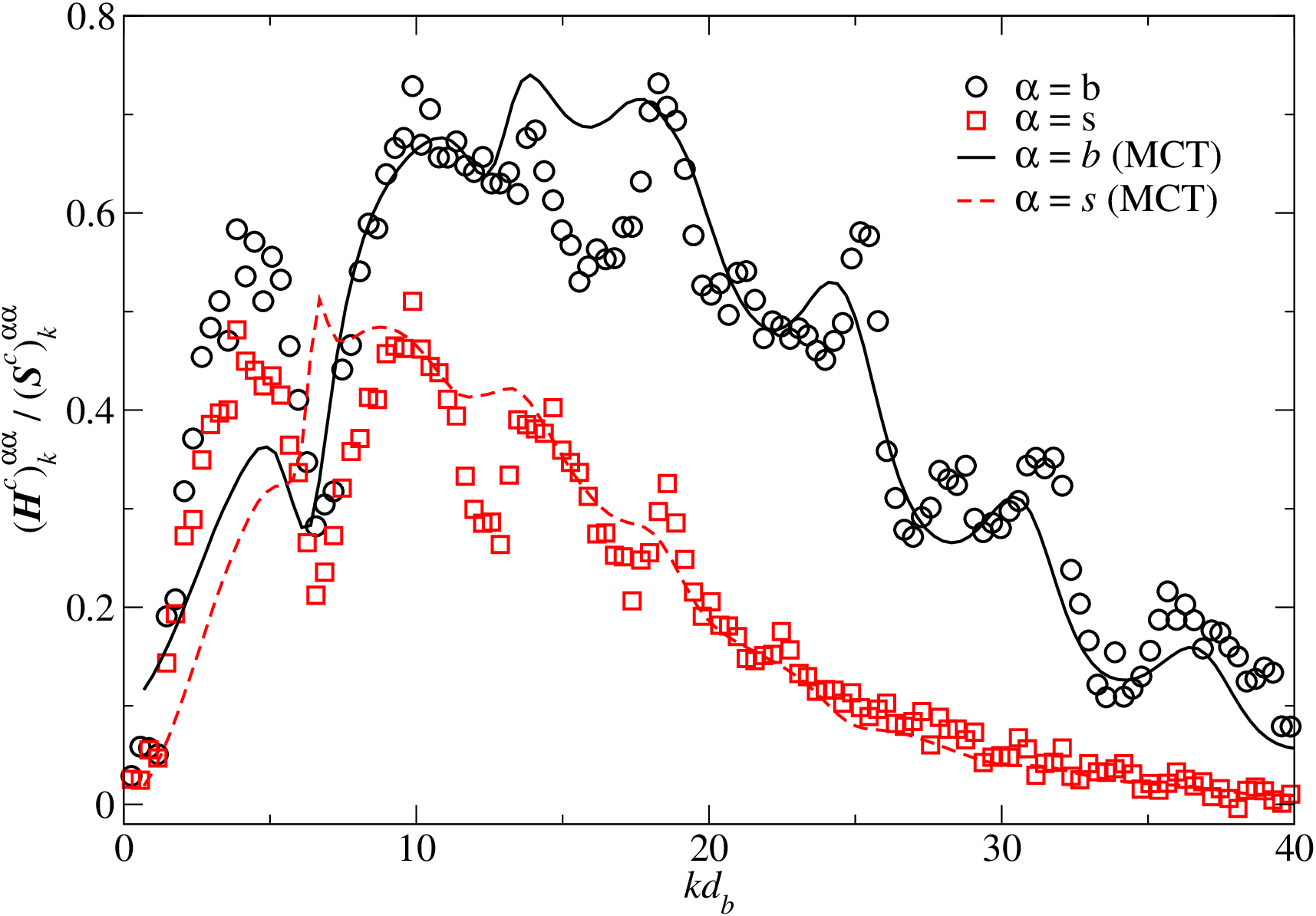}
\caption{(Color online) Critical amplitudes obtained by Eq.~(\ref{Yfunc}) for $\delta=1/3$ and $x_s=0.5$ with $k_0d_b=5.37$ from the simulated correlators at $\varphi =0.81$, $\delta = 1/3$ and $x_s=0.5$. Solid black and dashed red lines depict the MCT results obtained with the same simulated structure factors as used in Fig.~\ref{delta3-fq-big} and Fig.~\ref{delta3-fq_small.eps}.}
\label{hq-fig_3.0}
\end{figure}

Let us investigate the so-called critical amplitude as a further interesting static quantity. 
Fig.~\ref{hq-fig_1.4} shows approximate values for the normalized critical amplitudes
$(\boldsymbol{H}^c)_k^{\alpha\alpha}/(\boldsymbol{S}^c)_k^{\alpha\alpha}$ for both the big and the small disks
at $\delta=5/7$ and $x_s=0.5$, extracted from our
BD simulations. Corresponding results for $\delta=1/3$ are shown in Fig.~\ref{hq-fig_3.0}.
In order to determine the critical amplitudes from the BD simulation data one can define the function \cite{graeser2006}
\begin{equation}\label{Yfunc}
Y_{k}^{\alpha\beta} = \frac{ \sum_{j=1}^{n/2} \Phi_k^{\alpha\beta}(t_j)-\sum_{j=n/2+1}^{n} \Phi_k^{\alpha\beta}(t_j)}
  {\sum_{j=1}^{n/2} \Phi_{k_0}^{\alpha\beta}(t_j)-\sum_{j=n/2+1}^{n} \Phi_{k_0}^{\alpha\beta}(t_j)} =
  \frac{(\boldsymbol{H}^c)_k^{\alpha\beta}}{(\boldsymbol{H}^c)_{k_0}^{\alpha\beta}}
\end{equation} 
with $t_j$ chosen in the $\beta$-scaling regime which in our case is determined to $t_j \in [0.47449, 8.2882]$ for $\delta=5/7$
and $t_j \in [0.10334, 1.8051]$ for $\delta=1/3$.
The last equality follows from Eq.~(\ref{factorization}) and thus allows us to extract the critical amplitudes  $(\boldsymbol{H}^c)_k^{\alpha\beta}$ up to a factor $(\boldsymbol{H}^c)_{k_0}^{\alpha\beta}$. 
Beside the numerical uncertainty at low $kd_b$, the simulation data for $\delta=1/3$ are in a good agreement with our
corresponding results from MCT using BD-simulated structure factors as input. For the case $\delta=5/7$ we observe larger discrepancies for all $kd_b$.

\subsection{Dynamics}

\begin{figure}
\includegraphics[width=1\columnwidth]{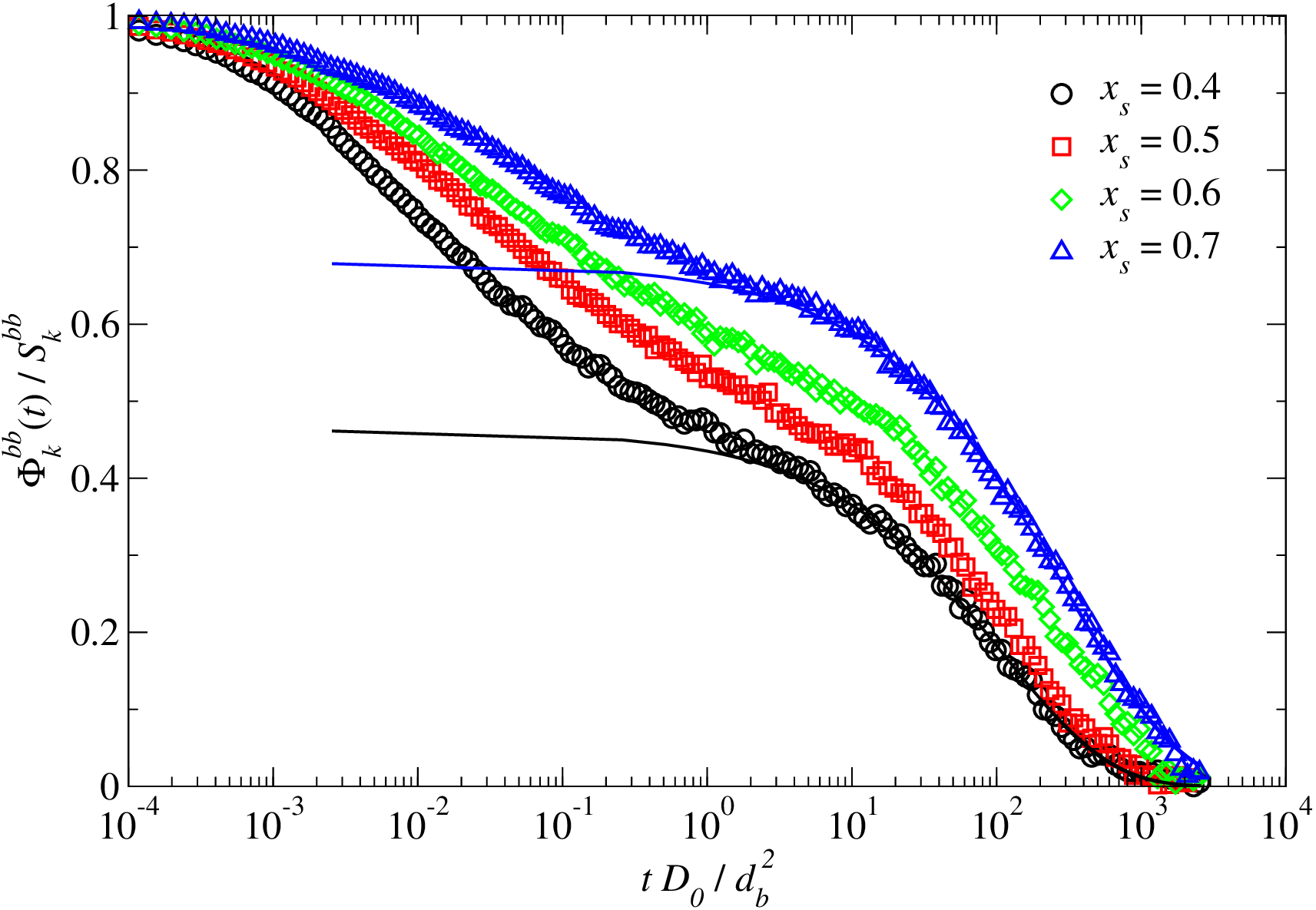}
\caption{(Color online)  Simulated normalized collective correlation functions of the big particles for the size ratio $\delta = 5/7$ at $\varphi = 0.79$ and $kd_b =8.5$ for varying $x_s \in \lbrace 0.4, 0.5, 0.6, 0.7 \rbrace$ as labeled in the legend. Solid lines show examples of Kohlrausch fits to the $x_s=0.4$ correlator (black) and the $x_s=0.7$ correlator (blue). }
\label{delta1.4-phi79-q6-big}
\end{figure}
\begin{figure}
\includegraphics[width=1\columnwidth]{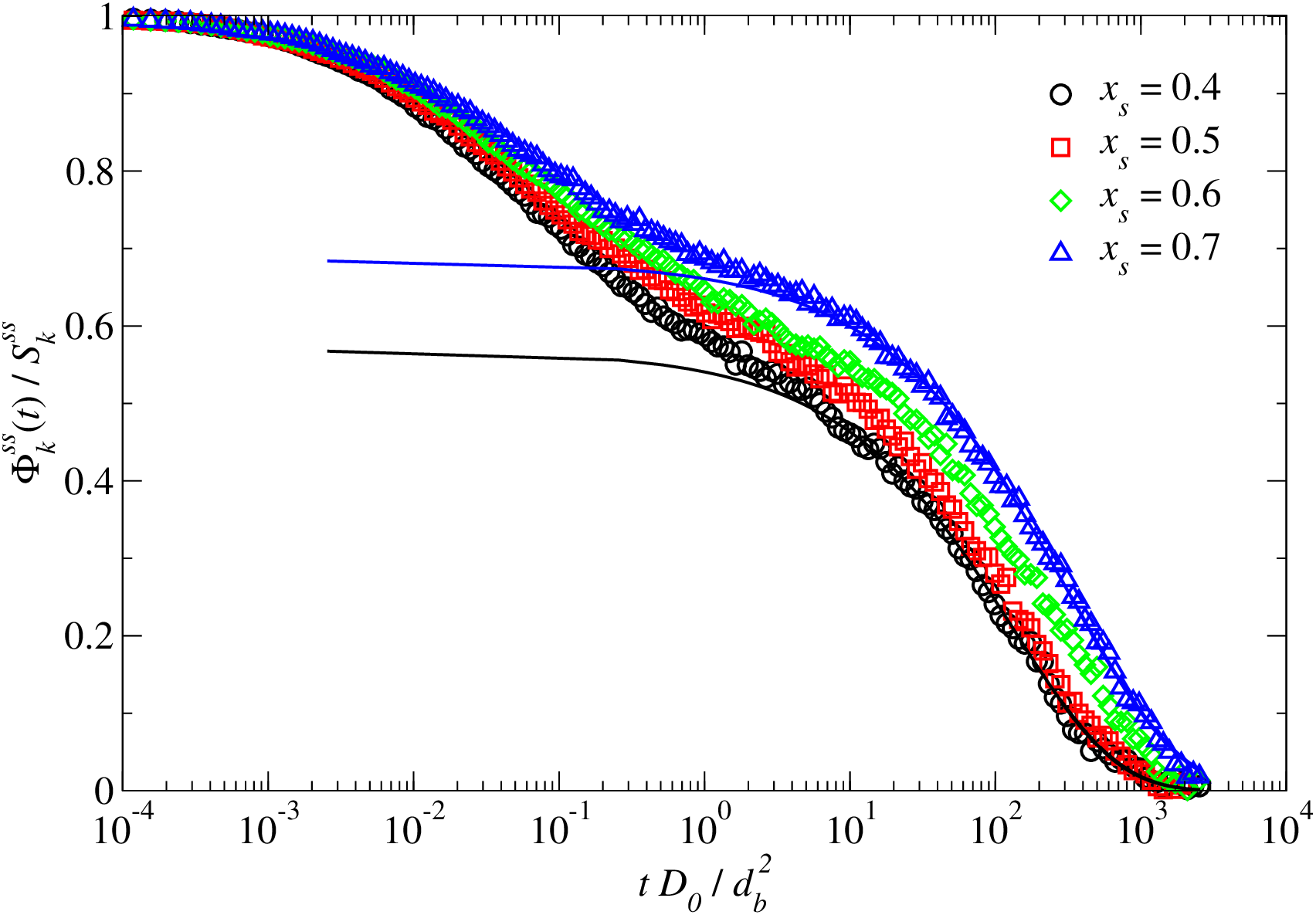}
\caption{(Color online) Simulated normalized collective correlation functions of the small particles for the size ratio $\delta = 5/7$ at $\varphi = 0.79$ and $kd_b=8.5$ for varying $x_s$. The color and symbol coding is the same as in Fig.~\ref{delta1.4-phi79-q6-big}. Solid lines show examples of Kohlrausch fits to the $x_s=0.4$ correlator (black) and the $x_s=0.7$ correlator (blue).}
\label{delta1.4-phi79-q6-small}
\end{figure}

In this section the most important quantities for MCT, the collective density correlators will be discussed.
Fig.~\ref{delta1.4-phi79-q6-big} shows normalized collective density correlators $\Phi_k^{bb}(t)/S_k^{bb}$ from our BD simulations for the big particles of binary hard disk mixtures in 2D at $\delta=5/7$, $\varphi=0.79$ and $kd_b=8.5$ for different concentrations ${x}_s$ of the smaller disks. Similar results for $\Phi_k^{ss}(t)/S_k^{ss}$ are shown in Fig.~\ref{delta1.4-phi79-q6-small}.
Focusing on the data for $\Phi_k^{bb}(t)/S_k^{bb}$ in Fig.~\ref{delta1.4-phi79-q6-big},
these data exhibit the same three mixing effects as the ones from MCT shown in Fig.~6 in Ref.~\cite{hajnal2009},
namely (iii) an increase in the plateau values accompanied by (iv) a slowing down of the relaxation towards these plateaus and (i) an additional slowing down of the $\alpha$-relaxation process upon increasing $x_s$.
However, there are some deviations: from  Fig.~5 in Ref.~\cite{hajnal2009} we would expect that the slowest $\alpha$-relaxation process occurs at $x_s\cong0.5$. The simulation data, however, exhibit the slowest $\alpha$-relaxation at the highest investigated value $x_s=0.7$. Unfortunately the simulation systems at higher $x_s$ are subject to crystallization which makes them unsuitable for the MCT comparison, rendering this region 'unaccessible' for the simulations. A possible source for the discrepancy could be, that in Ref.~\cite{hajnal2009} an approximate theory, the Percus-Yevick structure factor was used as input to the MCT calculations.
Furthermore, for $\delta=5/7$, MCT predicts $0.686<\varphi^c<0.6920$ for all $x_s$. The simulation data, however, imply $\varphi^c_{sim}\gtrsim0.79$ which means that MCT underestimates the critical packing fraction by about $15\%$. The Percus-Yevick approximation contributes to this underestimation but even with simulated structure factors as input, MCT underestimates the critical packing fraction.
In Fig.~\ref{delta1.4-phi79-q6-small} we basically find the same scenario as in Fig.~\ref{delta1.4-phi79-q6-big},
although a bit less pronounced, as expected from MCT \cite{hajnal_diss}.

\begin{figure}
\includegraphics[width=1\columnwidth]{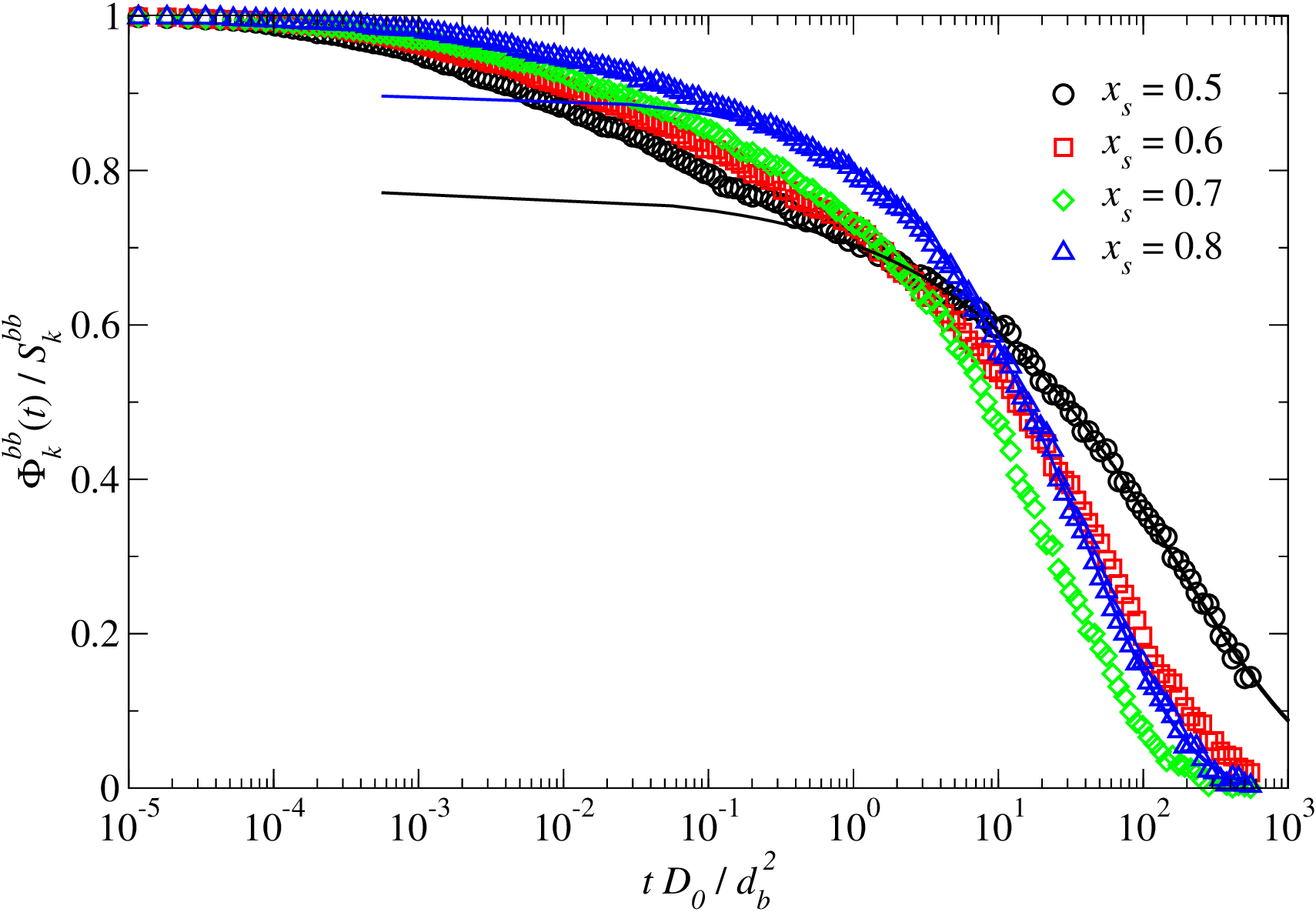}
\caption{(Color online) Simulated normalized collective correlation functions of the big particles for the size ratio $\delta = 1/3$ at $\varphi = 0.81$ and $kd_b=9.0$ for varying $x_s \in \lbrace 0.5, 0.6, 0.7, 0.8 \rbrace$ as labeled in the legend. Solid lines show exemplary Kohlrausch fits to the $x_s=0.5$ correlator (black) and the $x_s=0.8$ correlator (blue).}
\label{delta3-phi81-q3-big}
\end{figure}
\begin{figure}
\includegraphics[width=1\columnwidth]{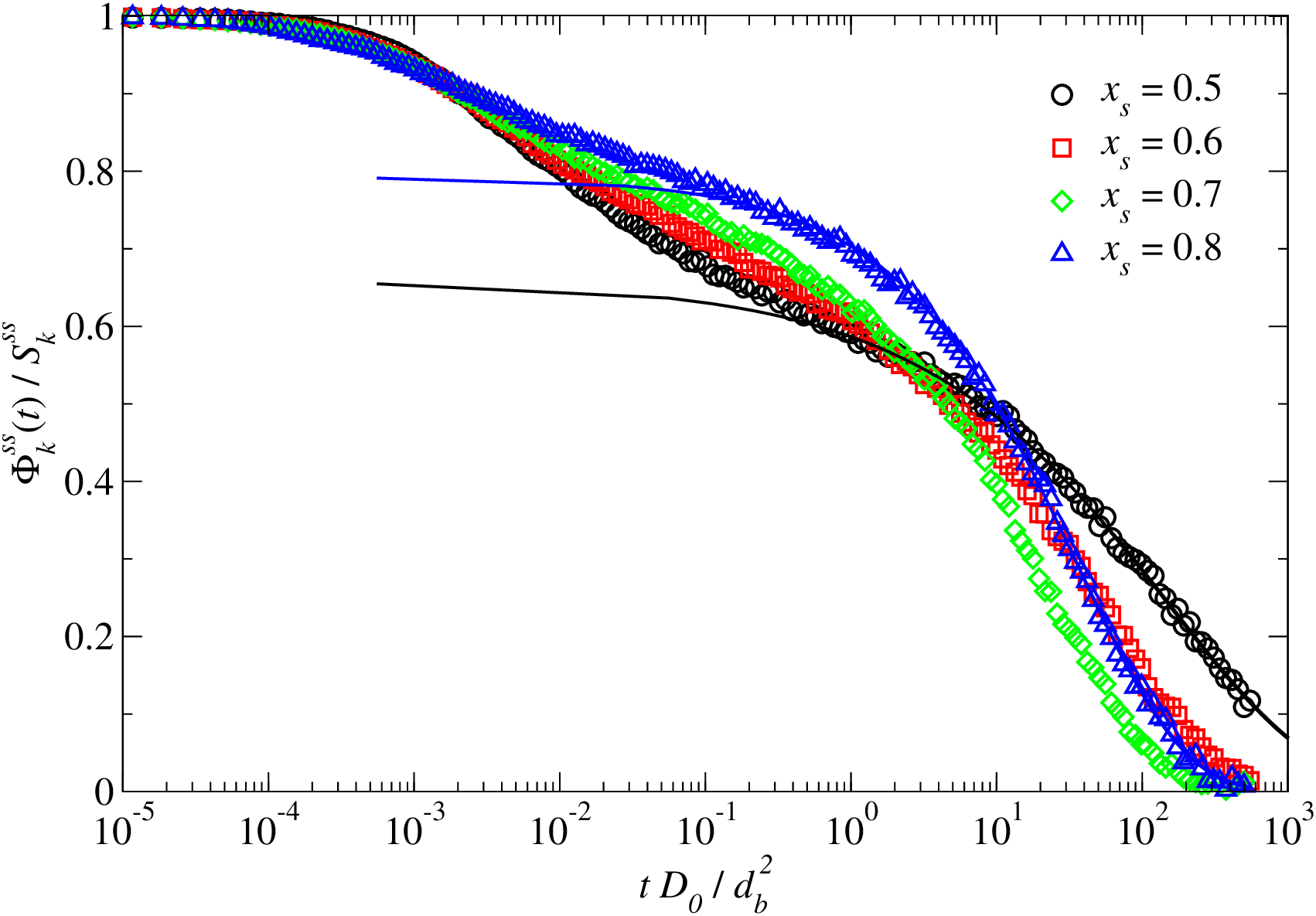}
\caption{(Color online)  Simulated normalized collective correlation functions of the small particles for the size ratio $\delta = 1/3$ at $\varphi = 0.81$ and $kd_b=9.0$ for varying $x_s$. The color and symbol coding is the same as in Fig.~\ref{delta3-phi81-q3-big}. Solid lines show exemplary Kohlrausch fits to the $x_s=0.5$ correlator (black) and the $x_s=0.8$ correlator (blue). }
\label{delta3-phi81-q3-small}
\end{figure}

Fig.~\ref{delta3-phi81-q3-big} shows the normalized collective density correlators $\Phi_k^{bb}(t)/S_k^{bb}$ from our BD simulations at $\delta=1/3$, $\varphi=0.81$ and $kd_b=9.0$ for different concentrations $x_s$ of the smaller disks. Similar results for $\Phi_k^{ss}(t)/S_k^{ss}$ are shown in Fig.~\ref{delta3-phi81-q3-small}.
First we have a closer look on the data for $\Phi_k^{bb}(t)/S_k^{bb}$ in Fig.~\ref{delta3-phi81-q3-big}.
On a qualitative level, the data for the three lowest values for $x_s$ are fully consistent with all MCT results in both Fig.~5 in Ref.~\cite{hajnal2009} and Fig.~7 in Ref.~\cite{hajnal2009}.
Upon increasing $x_s$, the simulation data exhibit the mixing effect (iii) an increase in the plateau values accompanied by (iv) a slowing down of the relaxation towards these plateaus.
In addition to that, increasing $x_s$ from $0.5$ to $0.7$ leads to (ii) a speeding up of the $\alpha$-relaxation. Thus, the three correlators corresponding to the lowest values for $x_s$ exhibit a pair-wise crossing.
A further increase in $x_s$ to $0.8$ leads again to a slowing down of the $\alpha$-relaxation process, although from the MCT results in Fig.~5 in Ref.~\cite{hajnal2009} we would expect the
fastest $\alpha$-relaxation at $x_s\cong0.85$. For the small particles we observe the similar albeit less pronounced effects, see Fig.~\ref{delta3-phi81-q3-small}.

The fact that the simulation data at $\delta=5/7$ and $\varphi=0.79$ and the ones at $\delta=1/3$ and $\varphi=0.81$ show very similar $\alpha$-relaxation times is on a qualitative level consistent with the MCT result $\varphi^c({x}_s,\delta=5/7)<\varphi^c({x}_s,\delta=1/3)$ for $0<x_s<1$.
Let us  conclude here with the statement that, at least on a qualitative level, the four mixing effects predicted by MCT for the binary hard disk model in 2D \cite{hajnal2009} are also observable in our computer simulations which supports the quality of MCT in 2D.

\section{The glass transition of a selected mixture}

In this section we select one of the systems ($\delta =5/7$, $x_s=0.5$) from the preceeding sections and perform a more profound MCT analysis including asymptotic checks and determining the MCT glass transition point. This will complete the MCT analysis of the system already discussed under shear in Ref.~\cite{henrich2009} with respect to the quiescent state.

In analogy to Eq.~(\ref{Yfunc}) it is possible to test another prediction of MCT.
In order to investigate the factorization theorem given by Eq.~(\ref{factorization}), we consider the function \cite{gleim2000}
\begin{equation}
\label{xeq}
 X_k^{\alpha\beta}(t)= \frac{\Phi_k^{\alpha\beta}(t)- \Phi_k^{\alpha\beta}(t')}{\Phi_k^{\alpha\beta}(t')-\Phi_k^{\alpha\beta}(t'')}
\end{equation}
with fixed times $t'<t''$ to be chosen appropriately from the $\beta$-scaling regime. Then
Eq.~(\ref{factorization}) predicts
\begin{equation}
 X_k^{\alpha\beta}(t)=\frac{G(t)-G(t')}{G(t')-G(t'')}+\mathcal{O}(|\sigma|)
\end{equation}
not to be dependent on wave number and particle index, to leading order in the separation parameter $\sigma$.
Thus it must be possible to fix two times $t'$ and $t''$ uniquely so that superimposing $X_k^{\alpha\beta}(t)$ for different $k$ yields a window in which all $X_k^{\alpha\beta}(t)$ collapse. An advantage of this procedure is that the critical amplitude drops out and doesn't need to be fitted.

\begin{figure}
\includegraphics[width=1\columnwidth]{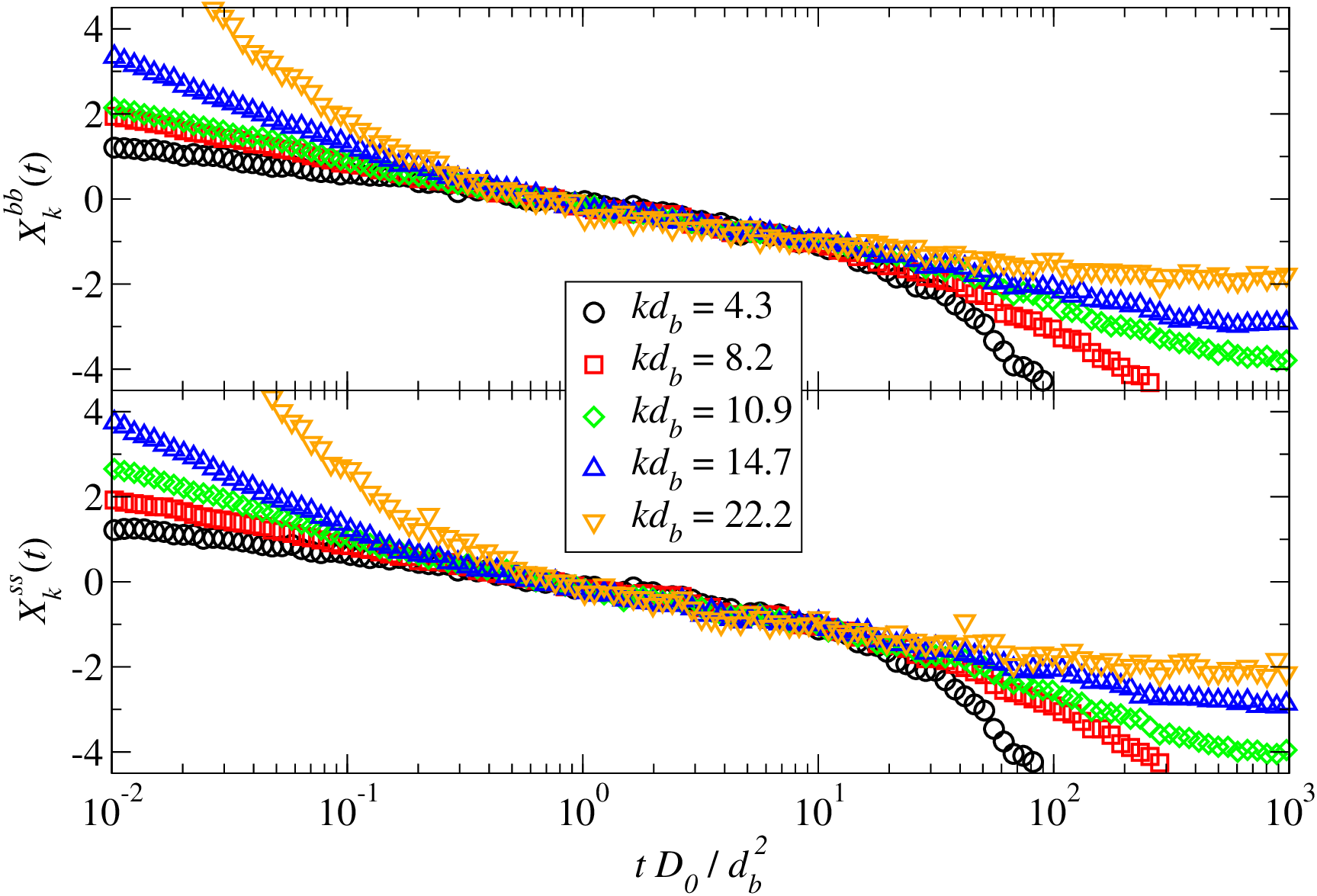}
\caption{(Color online) Functions $X^{\alpha\alpha}_k(t)$ calculated from Eq.~(\ref{xeq}) with the simulated correlators at $\varphi=0.79$, $\delta = 5/7$ and $x_s=0.5$
by fixing $t' D_0/d_b^2 = 0.7648$ and $t'' D_0/d_b^2 = 9.117$ for
the big and the small particles.}
\label{X-func-fig}
\end{figure}

Fig.~\ref{X-func-fig} shows our BD simulation results for $X_k^{\alpha\alpha}(t)$ at $\varphi=0.79$, $\delta = 5/7$, $x_s=0.5$, $t' D_0/d_b^2 = 0.7648$ and $t'' D_0/d_b^2 = 9.117$ for different wave numbers.
Indeed, within the numerical accuracy of our simulations, for both $\alpha=b$ and $\alpha=s$ the data for different $kd_b$ collapse onto each other within a time window of about two decades, similar to
previous results for binary Lennard-Jones mixtures in 3D \cite{gleim2000}.
A more sensitive test of MCT asymptotics is the so-called ordering rule.
As in the next-to-leading order corrections to the factorization theorem the same $k$-dependent correction amplitudes appear,
the deviations before the collapse regime must be in the same direction as after the collapse window. Hence correlators entering the collapse region in a certain order when numbered from top to bottom should leave the collapse window in exactly that ordering \cite{franosch1997}. Fig.~\ref{X-func-fig} is confirming that prediction.

\begin{figure}
\includegraphics[width=1\columnwidth]{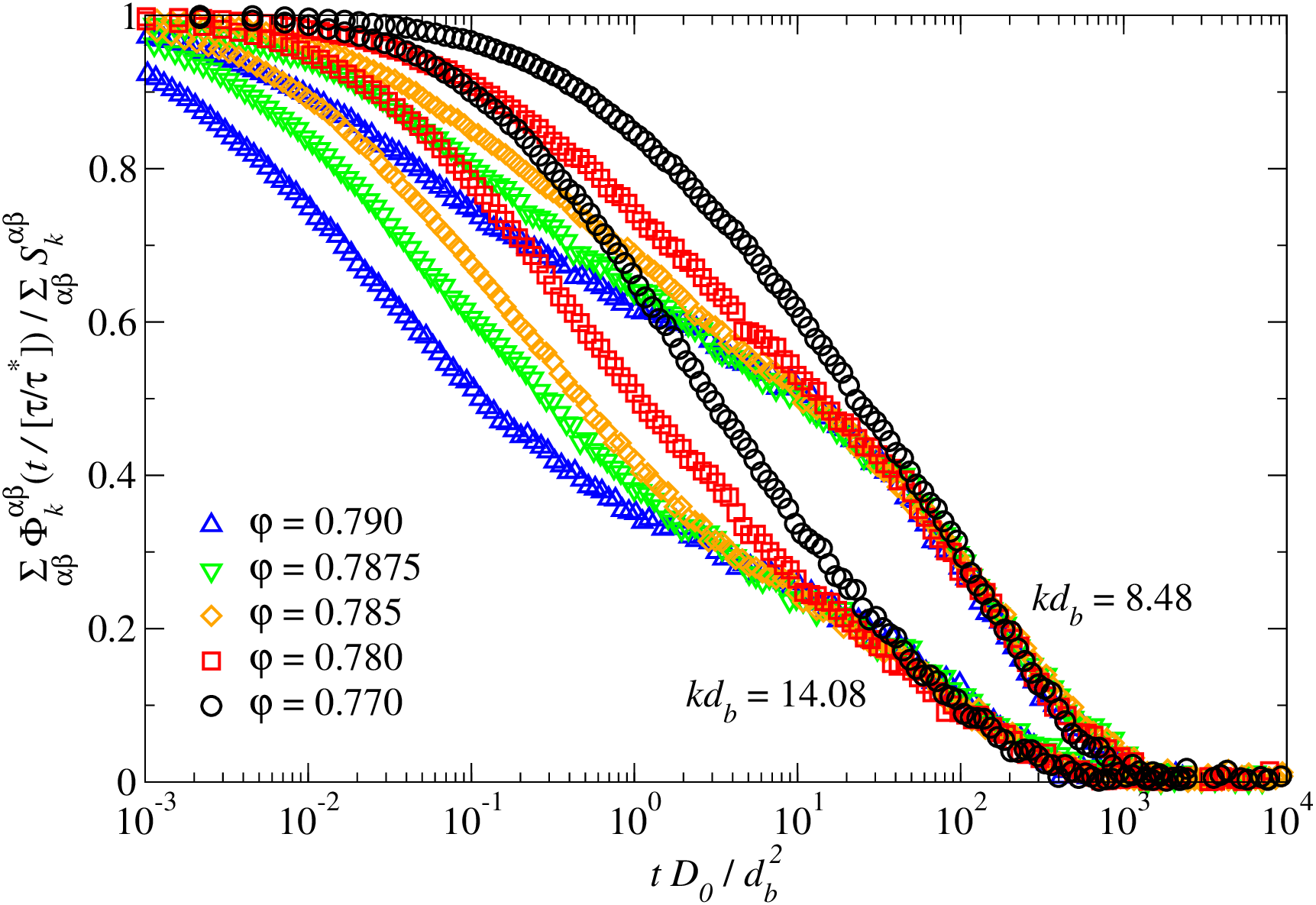}
\caption{(Color online) Rescaled correlators for $\varphi \le 0.79$, $\delta = 5/7$ and $x_s=0.5$ to collapse on one $\alpha$-master function at long times. The rescale times are independent on $kd_b$}
\label{rescaled}
\end{figure}

\begin{figure}
\includegraphics[width=1\columnwidth]{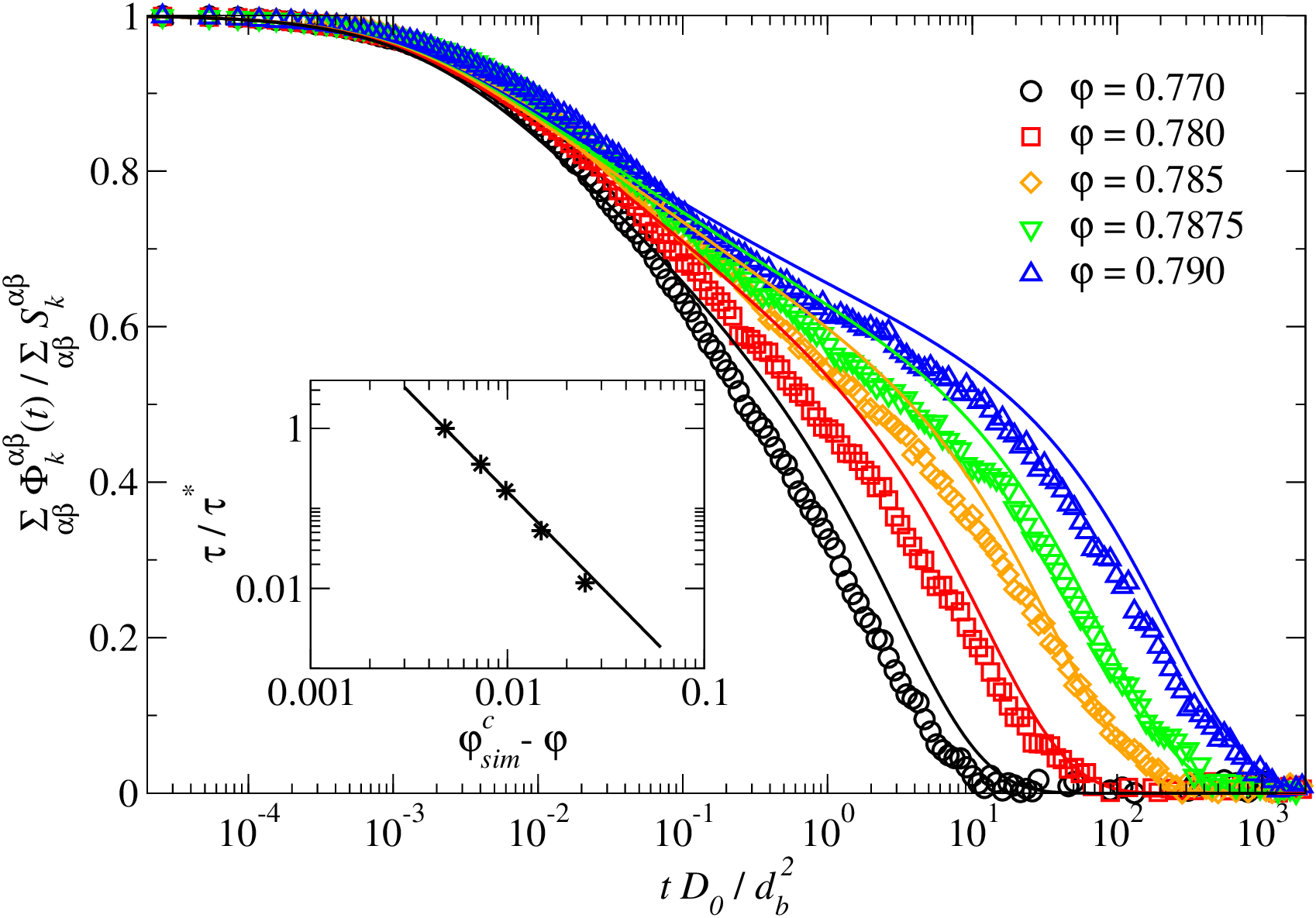}
\caption{(Color online) Normalized total collective correlators from BD simulations for $\delta = 5/7$, $x_s=0.5$ and $kd_b=8.5$.
The solid lines show corresponding MCT results where the packing fractions have been mapped according to Eq.~(\ref{mapping}).
The inset shows the $\alpha$-relaxation timescales and the fitted power law $\tau/\tau^* = A |\varphi-\varphi^c_{sim}|^{-\gamma}$ with
$\gamma=2.4969$ (fixed from MCT),
$A=1.633 \cdot 10^{-6}$ and $\varphi^c_{sim}=0.7948$. See text for details.}
\label{powerlaw_1.4}
\end{figure}

\begin{figure}
\includegraphics[width=1\columnwidth]{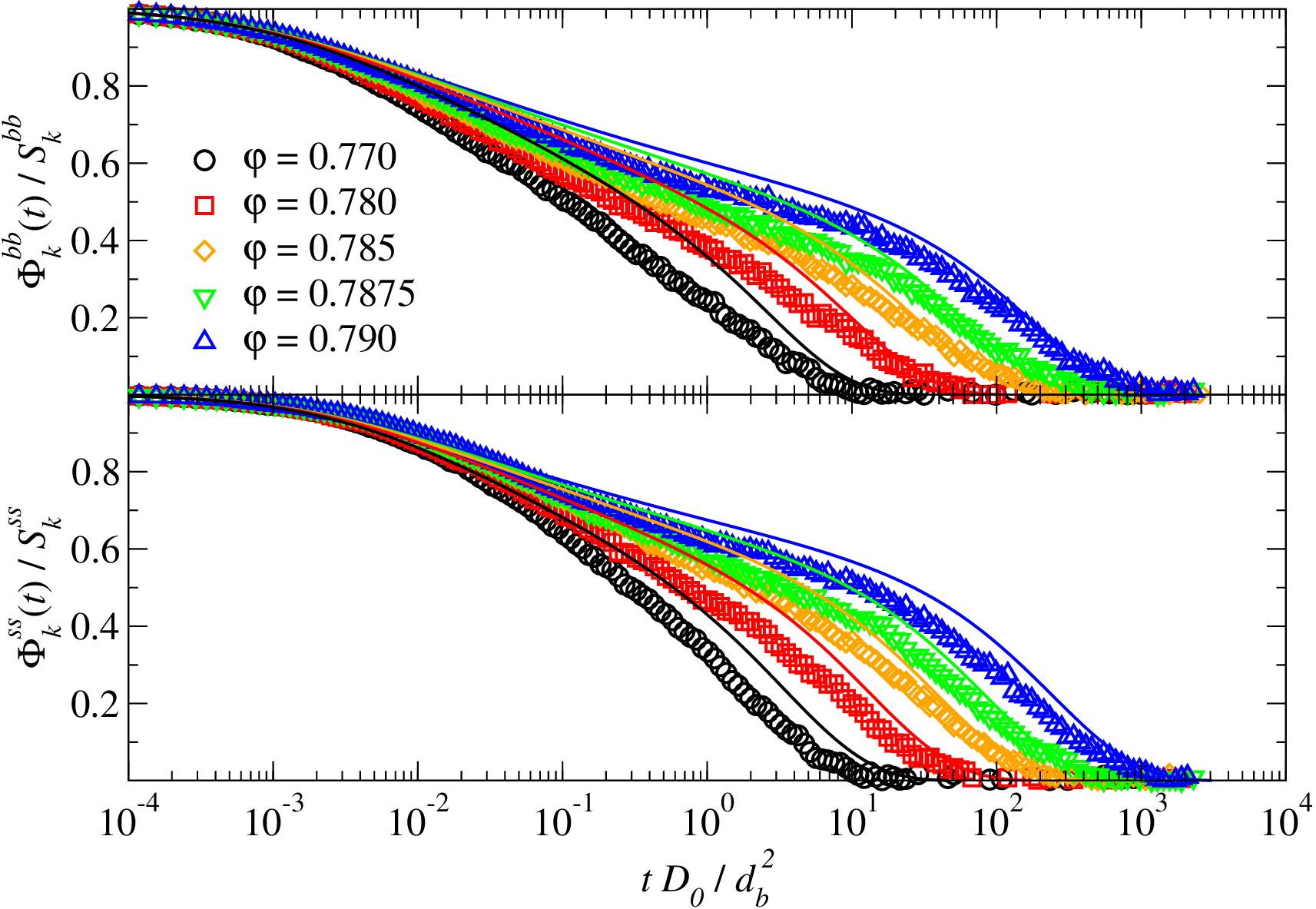}
\caption{(Color online) Normalized collective correlators from BD simulations for $\delta = 5/7$, $x_s=0.5$ and $kd_b=8.5$.
The solid lines show corresponding MCT results where the packing fractions have been mapped according to Eq.~(\ref{mapping}).}
\label{fit_8.5}
\end{figure}

\begin{figure}
\includegraphics[width=1\columnwidth]{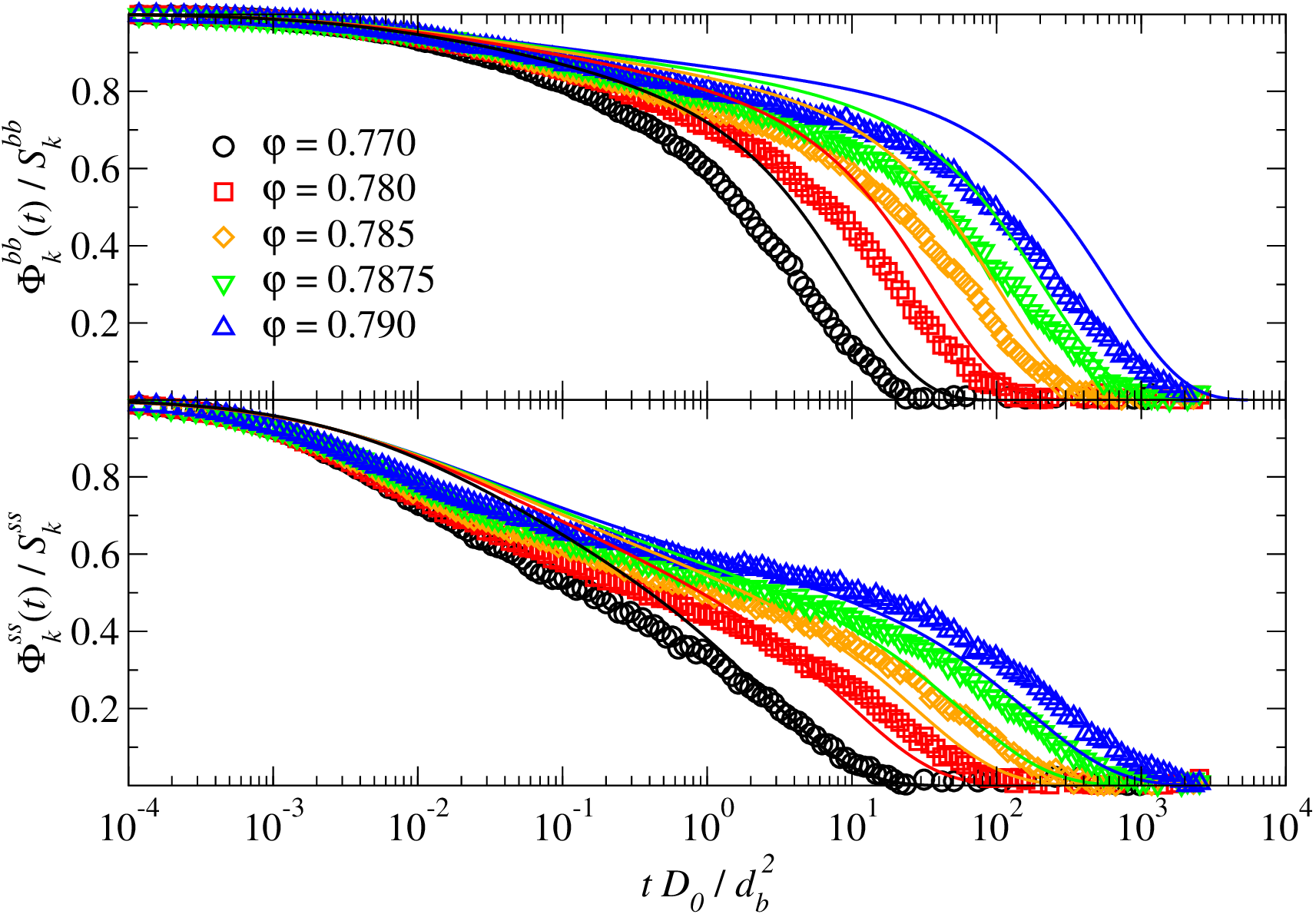}
\caption{(Color online) Normalized collective correlators from BD simulations for $\delta = 5/7$, $x_s=0.5$ and $kd_b=6.4$.
The solid lines show corresponding MCT results where the packing fractions have been mapped according to Eq.~(\ref{mapping}).}
\label{fit_6.4}
\end{figure}

\begin{figure}
\includegraphics[width=1\columnwidth]{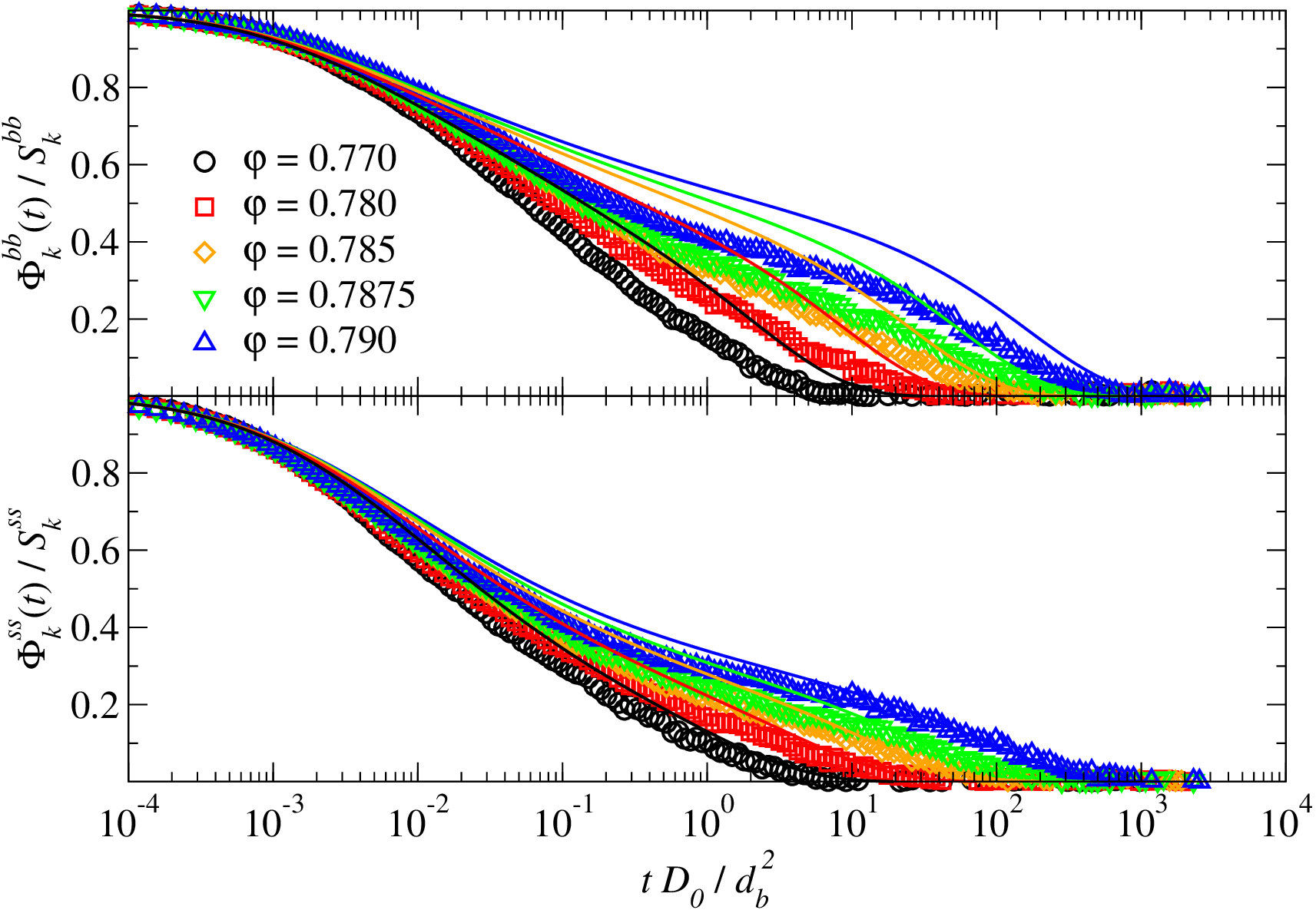}
\caption{(Color online) Normalized collective correlators from BD simulations for $\delta = 5/7$, $x_s=0.5$ and $kd_b=13.0$.
The solid lines show corresponding MCT results where the packing fractions have been mapped according to Eq.~(\ref{mapping}).}
\label{fit_13.0}
\end{figure}

Now, we test the validity of the $\alpha$-scaling law given by Eq.~\eqref{alpha_scaling_law}. According to
that equation, plotting the correlators as a function of $t/\tau$ makes the data collapse for long times on a master curve on
approaching $\varphi \to \varphi^c$ from the liquid.
To determine a dimensionless relaxation time $\tau/\tau^* \propto |\sigma|^{-\gamma}$ from the simulation, the correlators at
$kd_b \approx 8.5$ (at the structure factor peak of $\sum_{\alpha\beta}S_k^{\alpha\beta}$) corresponding to $\varphi<0.79$ have been shifted
along the $\log_{10}(t)$-axis
to coincide in the final decay with the one at the highest packing fraction $\varphi = 0.79$. The highest packing fraction defines
$\tau_{\varphi=0.79}/\tau^*=1$ and yields our best approximation for the $\alpha$-master function.
Checking that $\tau/\tau^*$ is independent on $kd_b$ we can validate the $\alpha$-scaling. We have chosen the structure factor peak
for the determination of $\tau/\tau^*$, as the strength of the $\alpha$-process is maximal here and thus a separation from the $\beta$-process can be achieved.
Fig.~\ref{rescaled} shows the exemplary result of the $\alpha$-scaling for two different wave numbers. The data clearly exhibit the two-step relaxation pattern of glass-forming liquids with increasingly stretched plateaus upon increasing the packing fraction. The shifted correlators approach an $\alpha$-master curve with the highest densities collapsing over almost three decades in time.

After having checked a few asymptotic results we now present full numerical MCT calculations. As a preliminary we use BD-simulated static structure factors as input for MCT to calculate the critical packing fraction $\varphi^c_{MCT}\cong0.6920$ and the exponent $\gamma=2.4969$. The obvious mismatch in $\varphi^c$ necessitates a comparison, at corresponding separation from the transition point. That entails matching the separation parameter $\sigma$ which is not easy to obtain from the simulation, but a peculiarity of our system helps us to circumvent the problem.
In Fig.~\ref{powerlaw_1.4} we show BD simulation results for the normalized total collective correlators $\sum_{\alpha\beta}\Phi_k^{\alpha\beta}(t)/\sum_{\alpha\beta}S_k^{\alpha\beta}$ at $\delta = 5/7$, $x_s=0.5$ and $kd_b=8.5$ for different packing fractions $\varphi$ within the liquid regime close to vitrification.
The inset in Fig.~\ref{powerlaw_1.4} shows our results for $\tau/\tau^*$ obtained from the shifting process seen in Fig.~\ref{rescaled}. The straight line shows the result from a power-law fit $\tau\sim|\varphi-\varphi^c_{sim}|^{-\gamma}$ with fixed $\gamma=2.4969$ which yields the extrapolated value $\varphi^c_{sim}\cong0.79481$ for the critical packing fraction which is approximately $15\%$ larger than the value predicted by MCT. For $|\varphi-\varphi^c_{sim}|\lesssim0.01$ the $\varphi$-dependence of $\tau/\tau^*$ is excellently described by the MCT exponent $\gamma$.
We can conclude: although MCT underestimates the critical packing fraction, it nevertheless describes very well the
$\varphi$-dependence of the $\alpha$-relaxation process in the liquid regime close to the glass transition,
which again supports the quality of MCT in 2D.

Using this information, we are able to present a quantitative comparison of time-dependent correlation functions from MCT to those from our BD simulations. For this purpose we have to take into account that MCT overestimates glass formation. It is well known that this results in predicting the glass transition at a too low critical packing fraction $\varphi^c$. Hence the relevant parameter when comparing MCT and simulation results is the separation parameter $\sigma$ which depends linearly on the distance parameter via $\sigma = C \varepsilon$ (see Section~\ref{sect_scaling}). As the constant $C$ is evaluated at the critical packing fraction with the corresponding structurefactor it is reasonable to assume that MCT doesn't yield the same prefactor as the simulation. For instance Flenner and Szamel found that both prefactors differ in a 3D binary Lennard Jones mixture \cite{FlennerSzamel2005}.
In order to construct a mapping of the packing fractions $\varphi_{sim}$ used in our BD simulation onto some appropriate ones $\varphi_{MCT}$ to be used for the corresponding MCT calculations we postulate that the separation parameters for both systems must be equal.  This leads us to the Ansatz
\begin{equation}
\label{mapping}
\varepsilon_{MCT}=(C_{sim}/C_{MCT}) \; \varepsilon_{sim} \equiv A \, \varepsilon_{sim}
\end{equation}
with some appropriately chosen constant $A$, which in our special case can be found empirically as $A\cong1$. 
With this, all input parameters for our MCT equations are uniquely determined. The solid lines in
Fig.~\ref{powerlaw_1.4} represent our MCT results for the normalized total collective correlation functions
corresponding to the shown BD simulation data in Fig.~\ref{powerlaw_1.4}.
Corresponding results for the normalized partial correlators $\Phi_k^{\alpha\alpha}(t)/S_k^{\alpha\alpha}$ are shown in Fig.~\ref{fit_8.5}.
We observe that for the chosen wave number MCT tends to underestimate the correlation functions in the transient time regime
$t D_0/d_b^2 \cong 2.5\cdot10^{-3}$ and overestimates the plateau values at times within the $\beta$-scaling regime.
Beside these quantitative deviations, MCT describes very well the qualitative $t$ and $\varphi$ dependences
of the BD-simulated correlation functions for $\varphi\geq0.78$. Note especially that
in this parameter regime MCT describes the final part of the $\alpha$-relaxation process also on the quantitative level correctly.

Of course, the grade of quantitative compliance of our BD simulation and MCT results is also dependent on the wave number.
We demonstrate this in Fig.~\ref{fit_6.4} for $\Phi_k^{\alpha\alpha}(t)/S_k^{\alpha\alpha}$ at $kd_b=6.4$ and in Fig.~\ref{fit_13.0} for
$\Phi_k^{\alpha\alpha}(t)/S_k^{\alpha\alpha}$ at $kd_b=13.0$. For instance, MCT strongly overestimates the plateau
values and also the $\alpha$-relaxation times for $\Phi_k^{bb}(t)/S_k^{bb}$ at both $kd_b=6.4$ and $kd_b=13.0$.
For $\Phi_k^{ss}(t)/S_k^{ss}$ at $kd_b=6.4$, on the other hand, MCT underestimates both the
plateaus and the $\alpha$-relaxation times. For $\Phi_k^{ss}(t)/S_k^{ss}$ at $kd_b=13.0$ MCT slightly overestimates the plateaus,
but excellently describes the $\alpha$-relaxation processes.
All these data indicate a connection between the (over)under-estimation of the plateaus and the (over)under-estimation
of the $\alpha$-relaxation times by MCT.

\begin{figure}
\includegraphics[width=1\columnwidth]{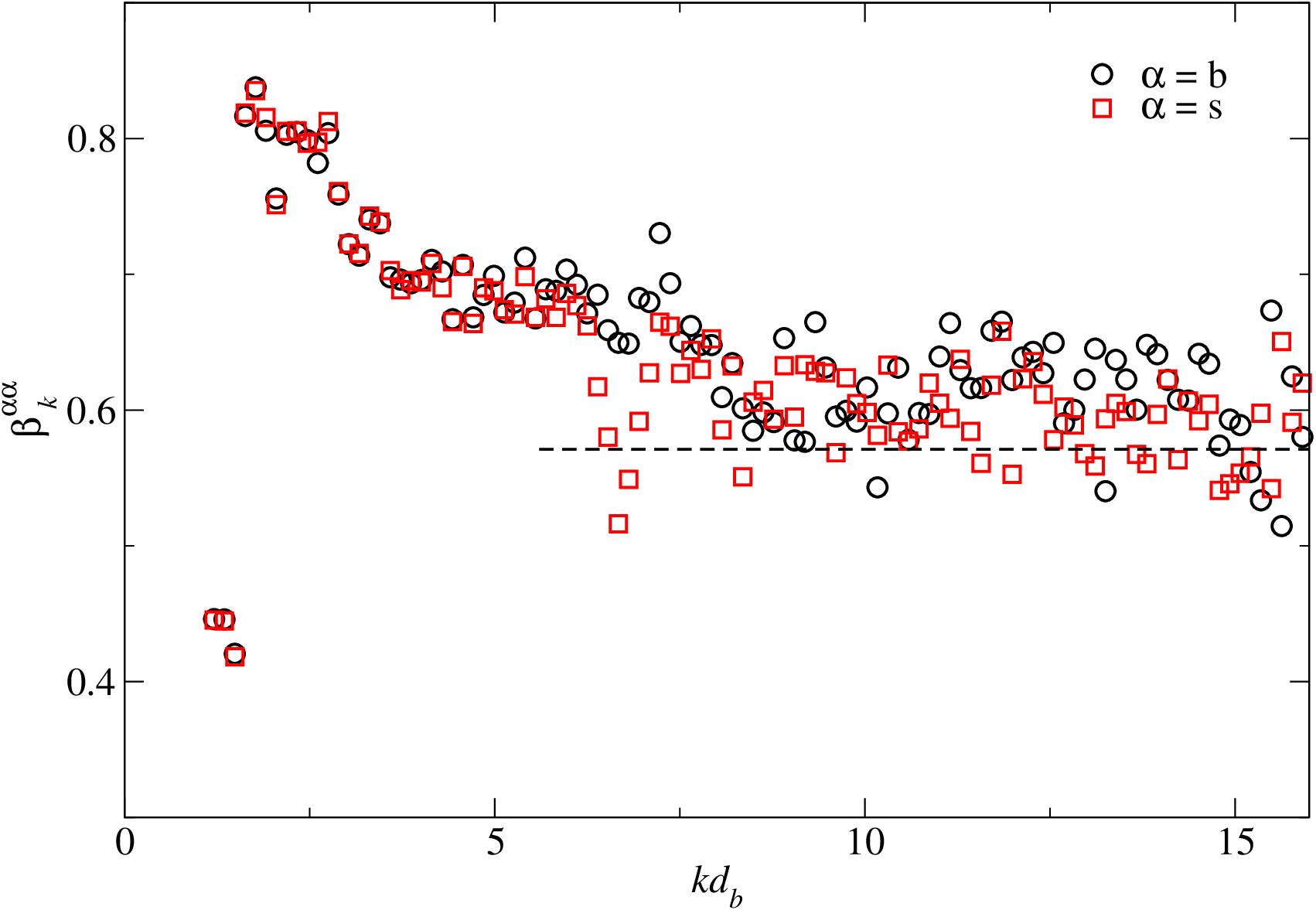}
\caption{(Color online) The Kohlrausch stretching exponent  $\beta_k^{\alpha \alpha}$ for the big (black circles)
and the small (red squares) particles at $\delta=5/7$, $x_s=0.5$ and $\varphi=0.79$. The MCT von Schweidler exponent $b=0.5571$ is indicated by the black dashed line. }
\label{betaq}
\end{figure}

For completeness we show the Kohlrausch stretching exponent in Fig.~\ref{betaq} for the small and big particles.
For large $kd_b$ the exponent converges as $\beta_k^{\alpha \alpha} \to 0.6 \pm 0.05$.
Unfortunately noisy data prevents a more precise determination of the high-$kd_b$ limit. Nevertheless the high-$kd_b$ limit is in good accordance
with MCT's von Schweidler exponent, Eq.~\eqref{vsmix}, calculated to $b=0.571$ with simulated structure factors.

\section{Beyond the MCT glass transition}

\begin{figure}
\includegraphics[width=1\columnwidth]{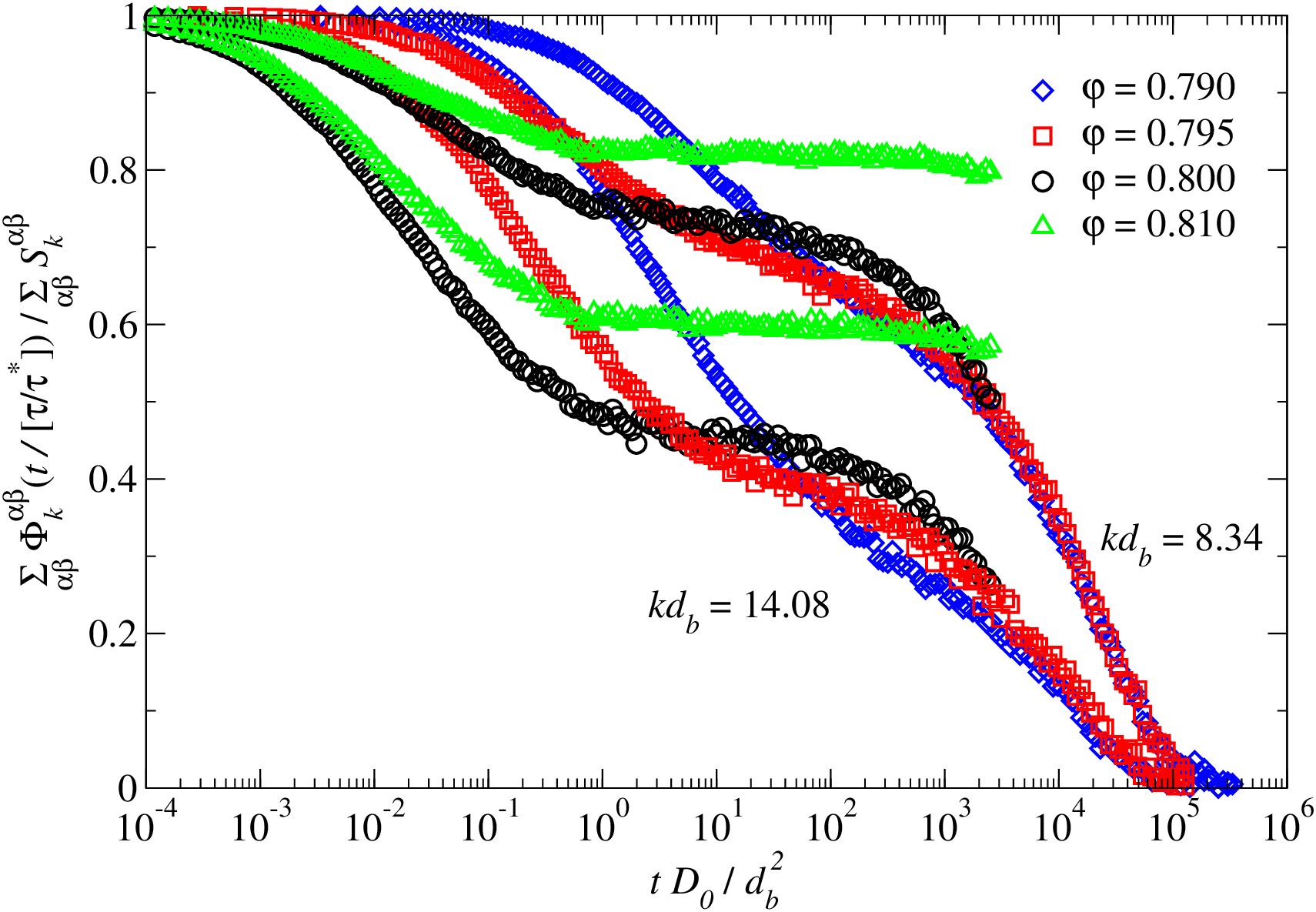}
\caption{(Color online) Three correlators above the MCT glass transition point at
$\delta=5/7$, $x_s=0.5$, $\varphi=0.81$, $\varphi=0.80$ and $\varphi=0.795$ together with the highest liquid density at $\varphi=0.79$. The correlators violate the scaling property shown in Fig.~\ref{rescaled}. The plateau values increase on going deeper into the glass}\label{nonscaling}
\end{figure}
Fig.~\ref{nonscaling} shows an example of three correlators with packing fractions above the MCT glass transition point.
Clearly one can see that the plateaus of the correlators are increasing with $\varphi$.
This leads to intersecting correlators and proves that the correlators do not collapse on a master function anymore
when they are rescaled by their relaxation times.
Furthermore the correlators above $\varphi \gtrsim \varphi^c \approx 0.7948$ show relaxation processes,
not captured in MCT which make the correlators beyond the MCT glass transition point decay to zero for long times.

The correlator at $\varphi=0.81$ does not show a relaxation within the time window accessible in our BD simulations. In the time window accessible by Newtonian dynamics a relaxation time could be extrapolated by fitting a Kohlrausch function to the coherent correlator at $kd_b = 8.4$.  These correlators were obtained by equilibrating 150 independent systems with Newtonian dynamics up to $\approx 20 \%$ of the structural relaxation time measured with the Kohlrausch fit. 
After having performed the Newtonian equilibration we do not see any dependence on the time origin for correlation functions in the time window accessible by Brownian dynamics. Especially the plateau values do not change any more. However the onset of a
final decay can be assumed, as the curves show a slight curvature for large times.

At this point we want to stress that even though the decay to zero is not included in MCT, MCT still gives an explanation for the rise of the NEPs
above the glass transition, see also Ref.~\cite{reinhardt2010}.

\section{Summary and conclusions}

We have performed BD simulations for binary mixtures of hard disks in 2D in order to test systematically
the predictions of MCT for this specific model system. Such a systematic test comparing collective density
correlation functions from both approaches has not been carried out before for a 2D model system.

Our main result is that MCT seems to be capable to capture many qualitative features of the glass transition
behavior of binary hard disk mixtures in 2D. Particularly, all four mixing effects predicted by MCT
for binary hard disks in 2D are indeed observable in our BD simulations. Furthermore, we have demonstrated
the validity of the factorization theorem given by Eq.~(\ref{factorization}) and the $\alpha$-scaling law given by 
Eqs.~(\ref{tau_alpha}) and (\ref{alpha_scaling_law}). We have found that
the MCT exponent $\gamma=2.4969$
describes excellently the dependence of the $\alpha$-relaxation process
on the packing fraction $\varphi$
in the liquid regime close to the glass transition.
Furthermore, MCT describes very well the qualitative $t$ and $\varphi$ dependences
of the BD-simulated collective correlation functions
close to the glass transition if one rescales the packing fraction according to  Eq.~(\ref{mapping}).
All these facts strongly support the quality of MCT in 2D. The range of validity of the asymptotic scaling
law is very similar to previous results for hard spheres in 3D.

Going beyond the extrapolated glass transition at $\varphi^c_{sim}\cong0.7948$ our BD simulation correlators loose the $\alpha$-scaling property which
results in intersecting correlators when they are rescaled by their relaxation times.

On the quantitative level, we have found some discrepancies between MCT and our BD simulation results.
For instance, MCT underestimates the absolute value for the critical packing fraction for vitrification by about $15\%$,
similar to previous results for hard spheres in 3D.
It seems to be possible to suppress the influence of this error
on time-dependent quantities by introducing a linear map according to  Eq.~(\ref{mapping}), at least for packing fractions
close to the glass transition. We have found quantitative deviations of the BD simulation results for the
NEPs and the critical amplitudes from the corresponding MCT results. For a better understanding of these deviations
it would be necessary to investigate the leading order corrections to the asymptotic scaling laws presented in Sect.~\ref{sect_scaling}.
This, however, would go far beyond the scope of our present study.

In the present paper we have verified mixing scenarios predicted by MCT for binary mixtures of hard disks in 2D.
Recently, mixing effects on the glass transition have also been investigated systematically in the framework of MCT for binary mixtures of dipolar point particles in 2D \cite{hajnal2010}.
As a project for the future it would be worth to also test the obtained MCT results systematically by BD simulations.

\acknowledgments
We thank Th. Voigtmann for discussions and hints concerning the treatment of binary mixtures.
Special thanks go to our supervisors M. Fuchs and R. Schilling who gave us the opportunity to conduct this project on our own.
\bibliography{lit}
\bibliographystyle{apsrev}

\end{document}